\documentclass[11pt,letter]{article}

\usepackage[affil-it]{authblk}
\usepackage{natbib}
\usepackage{amssymb,amsmath,amsfonts,amstext}
\usepackage{graphicx}
\usepackage{subfigure}
\usepackage{mathtools}
\usepackage{stmaryrd}
\usepackage{color}
\usepackage{fullpage}

\usepackage{amsthm}
\usepackage{esint}

\usepackage{float}
\usepackage{subfloat}
\usepackage{rotating}
\newcommand{\X}{\mathbf{X}}
\newcommand{\E}{\text{E}}

\newcommand{\Pb}{\mathbf{P}}
\newcommand{\B}{\mathbf{B}}
\newcommand{\p}{\boldsymbol \varphi}
\newcommand{\F}{\mathbf{F}}
\newcommand{\N}{\mathbf{N}}
\newcommand{\M}{\text{M}}

\graphicspath {
    {./}
    {./Figures/}
}

\title{\textbf{Discrete averaging relations for micro to macro transition}}

\author{Chenchen Liu, Celia Reina\footnote{Corrseponding author: creina@seas.upenn.edu}}
\affil{Department of Mechanical Engineering and Applied Mechanics, University of Pennsylvania, Philadelphia, PA 19104, USA}

\begin{document} 

\maketitle
\begin{abstract}
The well-known Hill's averaging theorems for stresses and strains as well as the so-called Hill-Mandel principle of macrohomogeneity are essential ingredients for the coupling and the consistency between the micro and macro scales in multiscale finite element procedures (FE$^2$). We show in this paper that these averaging relations hold exactly under standard finite element discretizations, even if the stress field is discontinuous across elements and the standard proofs based on the divergence theorem are no longer suitable. The discrete averaging results are derived  for the three classical types of boundary conditions (affine displacement, periodic and uniform traction boundary conditions) using the properties of the shape functions and the weak form of the microscopic equilibrium equations. The analytical proofs are further verified numerically through a simple finite element simulation of an irregular representative volume element undergoing large deformations. Furthermore, the proofs are extended to include the effects of body forces and inertia, and the results are consistent with those in the smooth continuum setting. This work provides a solid foundation to apply Hill's averaging relations in multiscale finite element methods without introducing an additional error in the scale transition due to the discretization.

\end{abstract}


\section{Introduction}
The vast majority of materials in nature as well as in engineering applications have underlying microstructures, and often, the length scale of the heterogeneites is much smaller than that of the system to be analyzed. In such cases, direct numerical simulations are typically prohibitive, and coarse-graining procedures have been developed to characterize the effective behavior of the material. A popular computational homogenization approach, that takes advantage of the separation of length scales, is the so-called FE$^2$ method. This numerical strategy considers a (coarse) finite element discretization for the macroscopic domain, and evaluates the effective behavior of the material at each quadrature point through a representative volume element (RVE), where the microstructure is resolved. This approach is capable of dealing with general geometries, materials, and loading conditions, and takes into consideration the evolution of the microstructure. It has been successfully applied to a wide range of problems, including composites \citep{feyel2000fe,terada2000simulation}, polycrystalline materials \citep{miehe1999computational,miehe2002homogenization,blanco2014variational}, elastic and plastic porous media \citep{smit1998prediction,reina2013micromechanical}, quasi-brittle separation laws  \citep{nguyen2011homogenization} and wave propagation in metamaterials \citep{pham2013transient}.


The coupling and energetic consistency between the two levels of representation in FE$^2$ methods is based on the seminal papers of Hill \citep{hill1963elastic,hill1967essential,hill1972constitutive} and subsequent developments by other authors \citep{mandel1972plasticite,willis1981variational,suquet1987elements,nemat1999averaging}. In particular, the often called Hill's theorems and Hill-Mandel principle of macrohomogeneity establish, in their original form, that the average strain and stress over the RVE are appropriate macroscopic quantities with which to describe the homogenized constitutive behavior, both in the linear and finite kinematic setting \citep{zohdi2008introduction,hori1999two}. These average quantities depend exclusively on the value of the corresponding microscopic object at the boundary of the RVE; and they thus enable the formulation of a boundary value problem (with either Neumann, Dirichlet or periodic boundary conditions) from which the effective behavior may be obtained.

To the best of the author's knowledge, the above micro-macro relations and variational reformulations \citep{miehe2002homogenization} and extensions to account for surfaces of discontinuities (\citet{nemat2013micromechanics}, Section 2.4), or body forces or inertial terms \citep{molinari2001micromechanical,ricker2009multiscale,reina2011multiscale,pham2013transient,de2015rve}, all rely on the strong form of the equilibrium equation and successive application of the divergence theorem.
However, these averaging relations are commonly used in finite element schemes, where the stress field may fail to be continuous, as is the case, for instance, for piecewise linear shape functions, and the balance equations are only satisfied in a weak sense. It is therefore, a priori, unclear whether Hill's relations hold exactly, or only approximately, in a discrete setting. This is an important issue for error estimations in multiscale finite element methods.

In this paper, the three fundamental averaging statements are shown to hold exactly for standard finite element discretizations: (i) the volume-averaged deformation gradient relation; (ii) the volume-averaged stress relation; and (iii) the energy average relation or so-called Hill-Mandel principle of macrohomogeneity. The proofs are conducted initially in the static setting with no body forces, and then extended to the more general case where body forces and/or inertial effects are present. In each case, the three classical types of boundary conditions are considered: affine displacement, periodic, and uniform traction boundary conditions. The statements are derived from the properties of the shape functions and the weak form of the momentum balance equations, and the use of the divergence theorem is limited to continuous fields such as the deformation mapping. In the interest of generality, the proofs are conducted in the finite kinematic setting. Finally, a finite element simulation over a highly irregular RVE with a coarse mesh is employed to numerically verify the discrete averaging relations. Simple extension and simple shear deformation modes are considered using displacement, tractions or periodic boundary conditions.

\section{Notation}
The analyses are based on the Lagrangian formulation of continuum mechanics. In this setting, the deformation of a material point $\mathbf{X}$ in the reference configuration $\Omega_0 \subset \mathbb{R}^3$ at time $t$ is characterized by the deformation mapping $\mathbf{x} = \boldsymbol \varphi (\mathbf{X},t)$. This mapping satisfies the momentum balance equations, which under sufficient smoothness, read
\begin{align} \label{Eq::EquilibriumEquations1}
& \nabla \cdot \Pb + \B=0, \qquad \text{in }  \Omega_0, \\ \label{Eq::EquilibriumEquations2}
& \boldsymbol \varphi = \bar{\boldsymbol \varphi}, \qquad \qquad \text{on}\ \partial \Omega_{0,1}, \\ \label{Eq::EquilibriumEquations3}
&  \mathbf{P}   \mathbf{N} =  \bar{\mathbf{T}}, \qquad \,\,\,\, \text{  on}\ \partial \Omega_{0,2}.
\end{align}
where $\mathbf{P}$ is the first Piola-Kirchhoff stress tensor, $\bar{\boldsymbol \varphi}$ and $\bar{\mathbf{T}}$ are the prescribed deformation mapping and traction respectively, and $\mathbf{N}$ is the outward unit normal to the domain in the undeformed configuration. For static loading, $\mathbf{B}=\mathbf{B}_0$ represents the body forces, whereas $\mathbf{B}$ also includes the inertial forces in the dynamic setting, i.e.~$\mathbf{B}=\mathbf{B}_0-\rho_0 \ddot{\p}$. Furthermore, the deformation gradient will be denoted by $\mathbf{F} = \nabla \boldsymbol \varphi$, where $\nabla$ represents the material gradient. As usual, it is required that $\partial \Omega_0 = \partial \Omega_{0,1} \cup \partial \Omega_{0,2}$, and $\partial \Omega_{0,1} \cap \partial \Omega_{0,2} = \emptyset$. In order to distinguish between the micro- and macro-variables, the superscript $\M$  will be employed for the latter, whereas no superscript will be used for the microscopic quantities.


The macroscopic fields will often result as the average of the corresponding microscopic objects. This average operation will be written as
\begin{equation}
\langle \cdot \rangle = \frac{1}{|\Omega_0|}\int_{\Omega_0} \cdot\ dV, 
\end{equation}
where $|\Omega_0|$ denotes the volume associated to $\Omega_0$ and $dV$ is the volume differential. Where needed, the surface differential will be denoted as $dS$.

Some derivations in the following sections will make use of standard indicial notation and Einstein summation convention. Lower case indices will then be used to refer to the deformed configuration and upper case indices for the reference configuration.

\section{Problem setting}
The multiscale finite element method FE$^2$ solves a boundary value problem for an RVE at each quadrature point of the macroscopic scale. Different types of boundary conditions can be imposed at the RVE in order to couple the micro- and macro-solution: affine displacement, periodic or traction boundary conditions. 

The linear displacement boundary conditions read 
\begin{equation}\label{Eq::BC_Hill}
\boldsymbol \varphi(\mathbf{X}) =\p^\M+ \F^\M \X,\qquad \text{on}\ \partial \Omega_0,
\end{equation}
where the macroscopic displacement field $\p^\M$ is often obviated in classical static analyses with divergence-free stresses, as a rigid body translation leaves the results unaltered, cf. \cite{nemat2013micromechanics} Chapter 1. However, in the presence of body forces or inertial, it is both, physically and mathematically meaningful to inform the RVE of the macroscopic translations, the rotations already being included in the deformation gradient \citep{ricker2009multiscale,reina2011multiscale,de2015rve}. 


Another boundary condition which is frequently employed is uniform traction boundary condition,
\begin{equation} \label{Eq:BC_T}
\mathbf{T}= \mathbf{P} ^{\M} \mathbf{N}, \qquad \text{on}\ \partial \Omega_{0}.
\end{equation}

It is well known that  \citep{suquet1987elements,peric2011micro} affine displacement boundary conditions result in stiffer solutions, whereas uniform traction boundary conditions are the most compliant. An intermediate behavior can be achieved with periodic boundary conditions
\begin{equation} \label{Eq::BC_UF}
\boldsymbol \varphi(\mathbf{X}) = \p^\M+\F^\M\mathbf{X}+ \tilde{\boldsymbol \varphi},\qquad \text{on}\ \partial \Omega_0,
\end{equation}
where $\tilde{\boldsymbol \varphi}$ stands for the fluctuation field, which is periodic along each pair of parallel sides (faces in three dimensions). Similar to the case with affine boundary conditions, the macroscopic translation is unnecessary for static problems with no body forces.




\subsection{ Finite element discretization and equilibrium equations}
Once a coupling strategy between the micro and macro scale is chosen, the multiscale FE$^2$ problem, obeying Eqs.~\eqref{Eq::EquilibriumEquations1}-\eqref{Eq::EquilibriumEquations3} at both scales, is resolved via a finite element discretization, not necessarily of the same type for the micro and macro problem. Here we consider conventional $C^0$ finite element discretizations \citep{hughes2012finite} for the RVE of the form
\begin{equation}
\boldsymbol \varphi^h (\X) = \sum_{a} \boldsymbol \varphi_a N_a(\X),
\end{equation} 
where $\{a\}$ represents the set of nodes, with associated degrees of freedom $\boldsymbol \varphi_a$, and $N_a$ are the corresponding shape functions, smooth within each element. These shape functions are required to have local support (each $N_a$ vanishes over any element not containing the node $a$) and to have the Kronecker delta property, $N_a(\X_b)=\delta_{ab}$. Furthermore, they shall satisfy 
the properties of partition of unity and linear field reproduction, i.e.,
\begin{align} \label{Eq:PartUnity}
&\sum_{a } N_a(\X) = 1,  \text{ and}\\ \label{Eq:Linear}
&\sum_{a } N_a(\X) X_{aJ} = X_J, 
\end{align}
which allow an exact representation of arbitrary rigid body motions and uniform deformations. These requirements for the shape functions will be essential for the derivations in Sections \ref{Sec:DiscreteTheorems_NoB} and \ref{Sec:DiscreteResults_B}.

\begin{figure}[htbp]
\begin{center}
    {\includegraphics[width=0.5\textwidth]{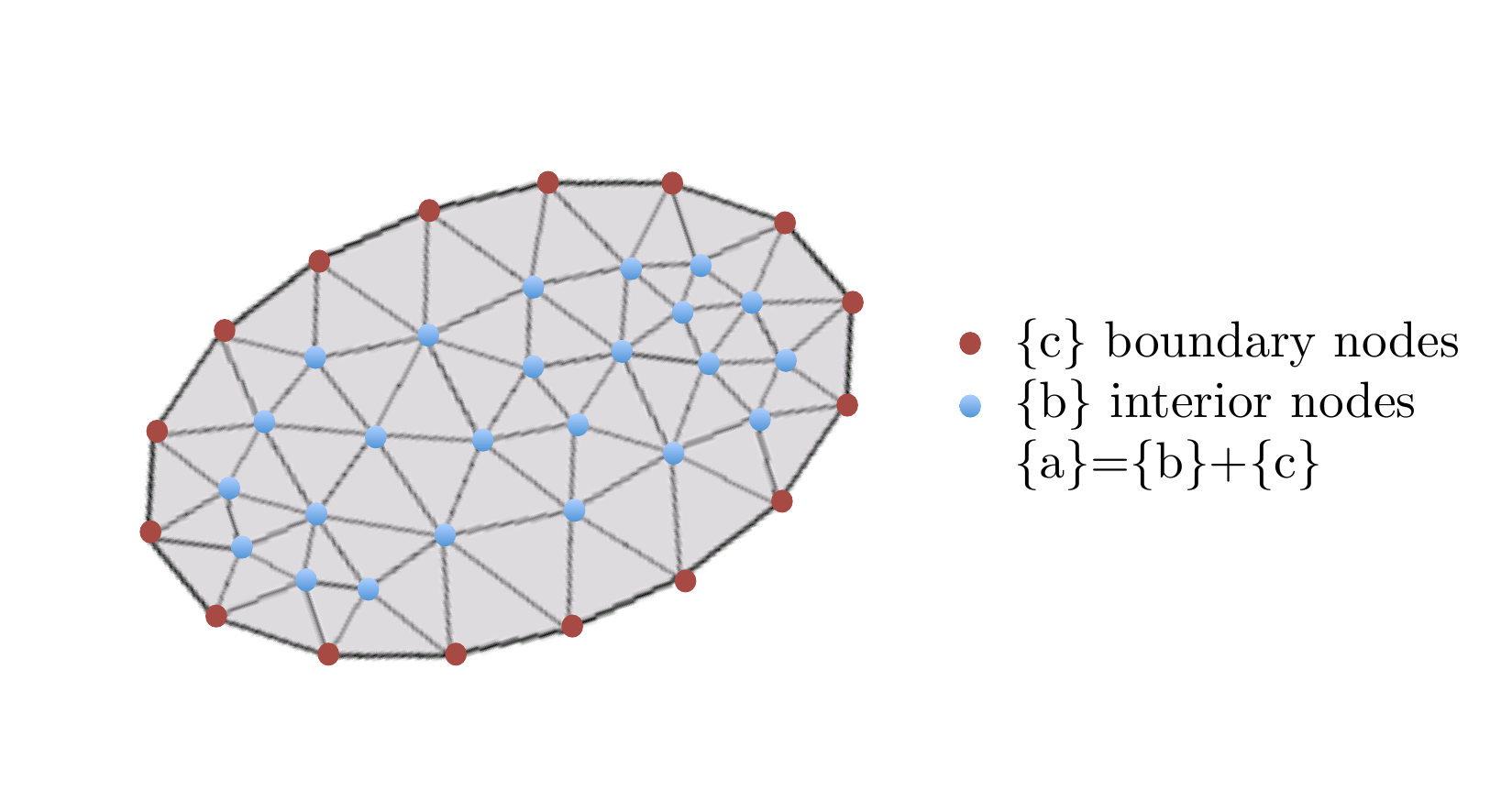}}
    \caption[]{Schematic representation of a finite element discretization, where the nodes are colored according to their location, in the interior or at the boundary of the domain.} \label{Fig:SetNodes}
\end{center}
\end{figure}

The weak form of the equilibrium equations associated the finite element discretization can then be readily obtained from the strong form, cf.~Eq.~\eqref{Eq::EquilibriumEquations1}. Towards that goal, it will result convenient to separate the node set $\{ a\}$ into interior nodes (\underline{b}ody nodes) $\{b\}$ and boundary nodes (at the \underline{c}ontour of the domain) $\{c\}$, cf.~Fig.~\ref{Fig:SetNodes}, and make use of the fact that the shape functions $N_b$ have a zero value at the boundary of the domain $\partial \Omega_0$ (by the local support property of the shape functions). Then, the weak form of the balance equations reads
\begin{equation} \label{Eq:EqWeak}
\int_{\Omega_0} \left(\nabla \cdot \Pb + \B\right)\cdot \delta \boldsymbol \varphi^h = 0 \rightarrow \sum_a \delta \varphi_{ai} \int_{\Omega_0} \left(P_{iJ} N_{a,J} - B_i N_a \right)\, dV -\sum_c \delta \varphi_{ci}\int_{\partial \Omega_0}P_{iJ}N_J N_c \, dS= 0.
\end{equation}
Equation \eqref{Eq:EqWeak} shall be satisfied for any admissible variation of the nodal positions $\delta \boldsymbol \varphi_a$, and, in particular, for the variation of any specific interior node $b$, 
while setting a null value for the variation of all the other nodes. 
Equivalently,
\begin{equation} \label{Eq:EqWeak_body}
\int_{\Omega_0} \left(P_{iJ} N_{b,J} - B_i N_b \right)\, dV = 0, \quad \text{for all the interior nodes }b.
\end{equation}
The variations of the nodes \{$c$\} depend on the boundary conditions used. For affine displacement boundary conditions, cf.~Eq.~\eqref{Eq::BC_Hill}, $\delta \boldsymbol \varphi_c$ =0, and therefore no additional equations follow. For periodic boundary conditions, cf.~Eq.~\eqref{Eq::BC_UF},  $\delta \boldsymbol \varphi_c= \delta \tilde{\boldsymbol \varphi}_c$ and the equilibrium equations for the boundary nodes read
\begin{equation} \label{Eq:EqWeak_periodic}
\sum_c \delta \tilde{\varphi}_{ci} \int_{\Omega_0} \left(P_{iJ} N_{c,J} - B_i N_c \right)\, dV = 0 \quad \text{ for all $\delta \tilde{\boldsymbol \varphi}_c$ periodic}.
\end{equation} 
In the absence of inertial and body forces, this equation is equivalent to anti-periodic boundary surface traction. 
Finally, for uniform traction boundary conditions, cf.~Eq.~\eqref{Eq:BC_T}, the weak form of the balance law for the boundary nodes is
\begin{equation} \label{Eq:EqWeak_traction}
\int_{\Omega_0} \left(P_{iJ} N_{c,J} - B_i N_c \right)\, dV - \int_{\partial \Omega_0}P_{iJ}^{\M}N_J N_c \, dS= 0.
\end{equation}
These equations may be readily simplified for the case with no body forces or inertia ($\B=0$), which is studied first.


\section{Discrete averaging results in the static case with no body forces}  \label{Sec:DiscreteTheorems_NoB}
The averaging statements for a representative volume element were initially developed for systems with negligible inertial and body forces and consist on the following three statements
\begin{itemize}
\item{} Averaging theorem for the deformation gradient: for any $\F$ compatible (i.e.~$\F =\nabla \boldsymbol \varphi$) and $\boldsymbol \varphi = \F^{\M} \X + \tilde{\boldsymbol \varphi}$ on $\partial \Omega_0$, with  $ \tilde{\boldsymbol \varphi}$ periodic or null, the macroscopic deformation gradient $\F^{\M}$ is equal to the average of its microscopic counterpart:  $\F^{\M}= \langle \F \rangle$.
\item{} Averaging theorem for the first Piola-Kirchhoff stress tensor: for any $\Pb$ in equilibrium (i.e.~$\nabla \cdot \Pb=0$) and $\mathbf{T}=\Pb^{\M} \N$ on $\partial \Omega_0$, the macroscopic stress tensor  $\Pb^{\M}$ is equal to the average of the microscopic analog: $\Pb^{\M} = \langle  \Pb \rangle$.
\item{} Hill-Mandel principle: for any $\F$ compatible and any $\Pb$ in equilibrium, not necessarily related to each other, and for any of the three standard types of boundary conditions (affine displacements, periodic or stress boundary conditions), the following equality holds: $\langle \Pb : \dot{\F} \rangle= \langle \Pb \rangle:\langle \dot{\F} \rangle$. In view of the previous two relations, this principle establishes the energy rate consistency between the micro and macro scale.
\end{itemize}


In this section, we show that these averaging results hold exactly under a finite element discretization. This implies that only the nodes at the boundary are required to satisfy the displacement boundary conditions, and the equilibrium equations, both, in the body and with the external tractions, are only satisfied weakly, cf.~Eqs.~\eqref{Eq:EqWeak_body}--\eqref{Eq:EqWeak_traction}. Furthermore, for the Hill-Mandel principle, the deformation gradient $\F$ and the stress tensor $\Pb$ will not be required to be related to each other, but they shall result from an identical finite element discretization (same set of nodes and associated shape functions).
The proofs are shown below for the three types of boundary conditions.

\subsection{Linear displacement boundary conditions}
The first case considered is that of affine displacement boundary conditions of an RVE, in accordance with the macroscopic deformation gradient $\mathbf{F}^{\M}$. In that case the boundary nodes {$\{c\}$} are required to satisfy
\begin{equation}\label{BC}
\varphi_{ci} = F^{\M}_{iQ} X_{cQ}.
\end{equation}

\subsubsection*{Averaging statement for the deformation gradient}
Equation \eqref{BC} is sufficient to show that the macroscopic deformation gradient tensor is the volume average of its microscopic analogue over the RVE. Since the displacement field is continuous over the domain and smooth in each element of the finite element discretization, the divergence theorem can be directly applied to $\nabla \boldsymbol \varphi^h$ over $\Omega_0$. Then,
\begin{equation} \label{Eq:Fave_1}
\begin{split}
\int_{\Omega_0}\varphi^h_{i,J} \, dV &= \int_{\partial \Omega_0}\varphi^h_i N_J \, dS =   \sum_a \varphi_{ai} \int_{\partial \Omega_0} N_a N_J \, dS= \sum_c \varphi_{ci} \int_{\partial \Omega_0} N_c N_J \, dS,
\end{split}
\end{equation}
where the sum  has been simplified to the boundary nodes, since the shape functions associated to the interior nodes, $N_b$, have a zero value at $\partial \Omega_0$. Next, we make use of the displacement boundary conditions on the node set {$\{c\}$}, cf.~Eq.~\eqref{BC}, and, again, use the fact that $N_b$ has a zero value at $\partial \Omega_0$
\begin{equation}
\begin{split}
\int_{\Omega_0}\varphi^h_{i,J} \, dV & = F_{iQ}^\M \sum_c  X_{cQ} \int_{\partial \Omega_0} N_c N_J \, dS = F_{iQ}^\M \sum_a  X_{aQ} \int_{\partial \Omega_0} N_a N_J \, dS.
\end{split}
\end{equation}
Finally, by the linear representation property of shape functions, cf.~Eq.~\eqref{Eq:Linear}, and the application of the divergence theorem, the sought-after result is obtained
\begin{equation}
\int_{\Omega_0}\varphi^h_{i,J} \, dV =F^\M_{iQ} \int_{\partial \Omega_0} X_Q N_J \, dS  =F^\M_{iQ} \int_{\Omega_0} X_{Q,J} \, dV= |\Omega_0| F^\M_{iJ},
\end{equation}
or equivalently,
\begin{equation}
\mathbf{F}^\M = \langle \mathbf{F} \rangle= \frac{1}{|\Omega_0|}\int_{\Omega_0} \mathbf{F} \, dV.
\end{equation}
In analogy with the continuum strain averaging result, the above proof is purely kinematical in nature and it is independent of the microscopic equilibrium equations.

\subsubsection*{Hill-Mandel principle}
Next, we proceed to prove the so-called Hill-Mandel principle for a compatible rate of deformation gradient $\dot{\F}= \nabla \dot {\boldsymbol \varphi}^h$, and a stress tensor $\Pb$ in equilibrium, i.e.~satisfying Eq.~\eqref{Eq:EqWeak_body} with $B_i=0$. The average microscopic (virtual) power can be written as
\begin{equation} \label{Eq:HillMandel_0}
\int_{\Omega_0} P_{iJ} \dot \varphi_{i,J}^h dV= \int_{\Omega_0} P_{iJ} \sum_a \dot \varphi_{ai} N_{a,J} \, dV=\sum_b \dot \varphi_{bi}  \left[\int_{\Omega_0} P_{iJ} N_{b,J} \, dV \right] + \int_{\Omega_0} P_{iJ} \sum_c \dot \varphi_{ci} N_{c,J} \, dV,
 \end{equation}
 where the sum over all nodes \{$a$\} has been divided into the sum over the interior nodes \{$b$\} and the boundary nodes \{$c$\}. The sum over \{$b$\} vanishes by the weak form of the equilibrium equations, cf.~Eq.~\eqref{Eq:EqWeak_body} with $B_i=0$; and the sum over \{$c$\} can be rewritten, applying the boundary conditions given by Eq.~\eqref{BC}, as
\begin{equation} \label{Eq:HillMandel1}
\int_{\Omega_0} P_{iJ} \dot \varphi_{i,J}^h dV =  \int_{\Omega_0} P_{iJ} \sum_c \dot F^\M_{iQ} X_{cQ} N_{c,J} \, dV =  \bigg[ \sum_c X_{cQ} \int_{\Omega_0} P_{iJ} N_{c,J} dV \bigg] \dot F^\M_{iQ}.
\end{equation}
The sum can then be extended to the set of all nodes by the equilibrium equations of the inner nodes, cf.~Eq.~\eqref{Eq:EqWeak_body} with $B_i=0$, from which it follows that
\begin{equation}
\int_{\Omega_0} P_{iJ} \dot F_{iJ} dV=  \bigg[  \int_{\Omega_0} P_{iJ} \sum_a X_{aQ} N_{a,J} dV\bigg] \dot F^\M_{iQ}.
 \end{equation}
Additionally, from the properties of the shape functions, cf.~Eq.~\eqref{Eq:Linear}, it is readily obtained that
 \begin{equation}\label{DeltaP}
\sum_a X_{aQ} N_{a,J} = X_{Q,J} = \delta_{QJ},
\end{equation}
and therefore
\begin{align}
\int_{\Omega_0} P_{iJ} \dot F_{iJ}\, dV = \left[  \int_{\Omega_0} P_{iJ} \, dV \right]\dot F^\M_{iJ}.
\end{align}
By making use of the previously derived relation for the deformation gradient, i.e. $F_{iJ}^\M=\langle F_{iJ} \rangle$, we have
\begin{align}
\langle \mathbf{P}: \dot\F \rangle = \langle \mathbf{P} \rangle : \langle \dot \F \rangle .
\end{align}
This completes the proof of Hill-Mandel principle under a finite element discretization with linear displacement boundary condition. 

\subsection{Periodic boundary condition} \label{Sec:PerBC_F}
The Dirichlet boundary conditions considered in the previous section may be relaxed, i.e.~softer response, by allowing a
periodic fluctuation field $\tilde{\boldsymbol \varphi}$. Under these considerations, the boundary nodes are required to satisfy the following relationship,
\begin{equation}\label{BCF}
\varphi_{ci} = F^{\M}_{iQ} X_{cQ} + \tilde\varphi_{ci}, \quad\text{with } \tilde\varphi_{ci}  \text{ periodic}.
\end{equation}

\subsubsection*{Averaging statement for the deformation gradient}
The proof of the average theorem for the deformation gradient follows in a very similar manner to the case with Dirichlet boundary conditions. Indeed, by making use of Eq.~\eqref{Eq:Fave_1} and the  displacement constraints on boundary nodes, cf.~Eq.~\eqref{BCF}, it follows that
%
\begin{equation}
\begin{split}
\int_{\Omega_0}\varphi^h_{i,J} \, dV & =  \sum_c \varphi_{ci} \int_{\partial \Omega_0} N_c N_J \, dS = F_{iQ}^\M \sum_c  X_{cQ} \int_{\partial \Omega_0} N_c N_J \, dS +\sum_c \tilde \varphi_{ci} \int_{\partial \Omega_0} N_c N_J \, dS.
\end{split}
\end{equation}
The last term on the right hand side of the equation vanishes as each pair of boundary edges have an identical displacement fluctuation and opposite outward normals. Then, by extending the sum over \{$c$\} to the full set of nodes (zero value of $N_b$ on $\partial \Omega_0$), making use of the linear field reproduction property of the shape functions, cf.~Eq.~\eqref{Eq:Linear}, and the divergence theorem, it follows that
\begin{equation}
\begin{split}
\int_{\Omega_0}\varphi^h_{i,J} \, dV &= F_{iQ}^\M \sum_a  X_{aQ} \int_{\partial \Omega_0} N_a N_J \, dS=F^\M_{iQ} \int_{\partial \Omega_0} X_Q N_J \, dS =F^\M_{iQ} \int_{\Omega_0} X_{Q,J} \, dV= |\Omega_0| F^\M_{iJ}.
\end{split}
\end{equation}
Thus, the classical averaging statement of the macroscopic deformation tensor is recovered, $\mathbf{F}^\M = \langle \mathbf{F} \rangle$, 
and is again independent of microscopic equilibrium equations.

\subsubsection*{Hill-Mandel principle}
Next, we proceed to prove Hill's energy consistency relation for a deformation rate $\dot{\F}$ compatible and deriving from a periodic mapping $\boldsymbol \varphi^h$, and for a stress tensor $\Pb$ satisfying the weak equilibrium equations given by Eqs.~\eqref{Eq:EqWeak_body} and \eqref{Eq:EqWeak_periodic} with $B_i=0$. Similarly to Eq.~\eqref{Eq:HillMandel_0}, the average of the microscopic energy rate obtained from the finite element solution is given by
\begin{equation}
\begin{split}
\int_{\Omega_0} P_{iJ} \dot \varphi_{i,J}^h \,dV&=\sum_b \left[\int_{\Omega_0} P_{iJ} N_{b,J} \, dV \right] \dot \varphi_{ib} + \int_{\Omega_0} P_{iJ} \sum_c \dot \varphi_{ci} N_{c,J} \, dV,
\end{split}
\end{equation}
where the set of nodes {$\{a\}$} has been divided into interior nodes {$\{b\}$} and nodes at the contour {$\{c\}$}. The term associated to the inner nodes vanished by the equilibrium equations. Then, by making use of the periodic boundary conditions, cf.~Eq.~\eqref{BCF}, and the anti-periodicity of $P_{iJ}N_{c,J}$, cf.~Eq.~\eqref{Eq:EqWeak_periodic}, one obtains
\begin{multline}
\int_{\Omega_0} P_{iJ} \dot \varphi_{i,J}^h \,dV =\int_{\Omega_0} P_{iJ} \sum_c \dot \varphi_{ci} N_{c,J} \, dV  =  \int_{\Omega_0} P_{iJ} \sum_c \left( \dot F^\M_{iQ} X_{cQ}+\dot{ \tilde{ \varphi}}_{ci} \right) N_{c,J} \, dV \\
 = \bigg[\int_{\Omega_0} P_{iJ} \sum_c X_{cQ} N_{c,J}\, dV \bigg] \dot F^\M_{iQ} +\sum_c \dot{\tilde{\varphi}}_{ci} \left[ \int_{\Omega_0} P_{iJ}  N_{c,J}  \, dV\right] = \bigg[\int_{\Omega_0} P_{iJ} \sum_c X_{cQ} N_{c,J}\, dV \bigg] \dot F^\M_{iQ} .
\end{multline}
The proof then follows in an identical manner to the case with affine boundary conditions, see derivations after Eq.~\eqref{Eq:HillMandel1}, leading as well to  $\langle \mathbf{P}: \dot\F \rangle = \langle \mathbf{P} \rangle : \langle \dot \F \rangle  $. This completes the proof of Hill-Mandel principle in discrete setting with periodic boundary conditions. 

\subsection{Uniform traction boundary conditions} \label{Sec:UniformTraction_noB}
The third and last case considered is that of uniform traction boundary conditions, given by the macroscopic stress tensor $\mathbf{P} ^{\M}$,
\begin{equation} \label{Eq:BC_UT}
\mathbf{T} = \mathbf{P} ^{\M} \N.
\end{equation}

\subsubsection*{Averaging statement for the first Piola-Kirchhoff stress tensor}
We proceed to show the volume-averaged relation of the stress tensor, under the boundary conditions given by Eq.~\eqref{Eq:BC_UT}. The volume integral of the microscopic stress over the RVE domain is
\begin{equation}
\int_{\Omega_0}P_{iJ} \, dV =\int_{\Omega_0}P_{iQ}\delta_{QJ} \, dV = \int_{\Omega_0} P_{iQ}\sum_a X_{aJ} N_{a,Q}\, dV= \sum_c X_{cJ}\int_{\Omega_0} P_{iQ} N_{c,Q}\, dV,
\end{equation}
where we have used the property of the shape functions given by Eq.~\eqref{DeltaP} and the equilibrium equations for interior nodes $\{b\}$, cf.~Eq.~\eqref{Eq:EqWeak_body} with $B_i=0$. Then, by substituting the weak form of the traction boundary condition, cf.~Eq.~\eqref{Eq:EqWeak_traction} with $B_i=0$, one obtains
\begin{equation}
\int_{\Omega_0}P_{iJ} \, dV = \sum_c X_{cJ}\int_{\partial \Omega_0} P_{iQ}^\M N_Q N_{c} \,dS= P_{iQ}^\M \sum_c X_{cJ}\int_{\partial \Omega_0} N_Q N_{c} \,dS.
\end{equation}
We may then extend the sum to all nodes $\{a\}$, as the shape functions associated to nodes $\{b\}$ are zero at $\partial \Omega_0$. This results in
\begin{equation}
\begin{split}
\int_{\Omega_0}P_{iJ} \, dV &= P^\M_{iQ} \sum_a X_{aJ}  \int_{ \partial \Omega_0}  N_Q N_{a} \, dS = P^\M_{iQ} \int_{ \partial \Omega_0} N_Q X_J= P^\M_{iQ} \int_{\Omega_0}  \delta_{JQ}\, dV = P^\M_{iJ} |\Omega_0|\\
\end{split}
\end{equation}
where we have further used the exact linear representation property of the shape functions, cf.~Eq.~\eqref{Eq:Linear}, and applied the divergence theorem.
The well-known averaging theorem is then recovered exactly
\begin{equation}
\mathbf{P}^\M = \langle \mathbf{P} \rangle = \frac{1}{|\Omega_0|}\int_{\Omega_0} \mathbf{P} \, dV,
\end{equation}
and follows from the equilibrium equations, in this case, in weak form.

\subsubsection*{Hill-Mandel principle}
Finally, we derive the Hill-Mandel principle for the finite element problem with stress boundary conditions. Similar to the previous section, we make use of the governing equations corresponding to the node sets $\{b\}$ and $\{c\}$, cf. Eqs.~\eqref{Eq:EqWeak_body} and \eqref{Eq:EqWeak_traction} with $B_i=0$, respectively, and obtain
\begin{equation}
\begin{split}
\int_{\Omega_0}P_{iJ} \dot \varphi_{i,J}^h \, dV&=\int_{\Omega_0}P_{iJ}\sum_a \dot \varphi_{ai}N_{a,J} \, dV=\sum_b \dot \varphi_{bi}\int_{\Omega_0}P_{iJ}N_{b,J} \, dV+\sum_c \dot \varphi_{ci}\int_{\Omega_0}P_{iJ}N_{c,J} \, dV \\
&=\sum_c \dot \varphi_{ci} \int_{\partial \Omega_0} P_{iJ}^\M N_J N_{c}\,dS=P_{iJ}^\M \sum_c \dot \varphi_{ci} \int_{\partial \Omega_0}  N_J N_{c}\,dS.\\
\end{split}
\end{equation}
The shape functions associated to nodes $\{b\}$ are zero at $\partial \Omega_0$ and the last sum of the above equation may thus be extended to the full set of nodes $\{a\}$. It then follows that 
\begin{equation}
\begin{split}
\int_{\Omega_0}P_{iJ} \dot \varphi_{i,J}^h \, dV&=P_{iJ}^\M \sum_a \dot \varphi_{ai} \int_{\partial \Omega_0}  N_J N_{a}\,dS=P_{iJ}^\M \sum_a \dot \varphi_{ai} \int_{\Omega_0}  N_{a,J}\,dV= P_{iJ}^\M \int_{\Omega_0} \dot \varphi_{i,J}^h \, dV,\\
\end{split}
\end{equation}
where the divergence theorem has been applied.
By substituting the previously derived relation $ P_{iJ}^\M= \langle P_{iJ} \rangle $, we obtain the sought-after result
\begin{align}
 \langle \mathbf{P}: \dot\F \rangle = \langle \mathbf{P} \rangle : \langle \dot \F \rangle .
\end{align}

\section{Discrete averaging results in the presence of body forces or inertia} \label{Sec:DiscreteResults_B}
In this section, we extend the previous results to account for the presence of body forces or inertial terms, which are both considered in the term $\B$. The boundary value problem for the RVE with periodic displacement boundary conditions and traction boundary conditions are considered separately, as well. The case with affine Dirichlet boundary conditions is omitted here, as its derivation is analogous to that of the periodic scenario.

\subsection{Periodic boundary condition}
In order to consider the work performed by the inertial and body forces, the boundary conditions on the microscopic domain read, for the parodic case, as
\begin{equation}\label{BCC}
\varphi_{ci} = \varphi^{\M}_i + F^{\M}_{iQ} X_{cQ}+\tilde\varphi_{ci}\, .
\end{equation}
where $ \boldsymbol \varphi^\M$ informs the RVE of the macroscopic translation.

\subsubsection*{Averaging statement for the deformation gradient}
As mentioned through the narrative, the averaging results for the deformation gradient are purely kinematical in nature and independent of the equilibrium equations. They thus hold as well in the presence of body forces or inertial terms. Note that the term $\boldsymbol \varphi^{\M}$ has an identical treatment to $\tilde{\boldsymbol \varphi}$ in the derivations and therefore, the proof remains the same as that of Section \ref{Sec:PerBC_F}. 

\subsubsection*{Hill-Mandel principle}\label{HM}

A micro-macro relation analogous to the classical Hill-Mandel principle has been obtained for non-zero value of the body forces or inertia \citep{molinari2001micromechanical,ricker2009multiscale,reina2011multiscale,pham2013transient,de2015rve}. It reads
\begin{equation} \label{Eq:HillMandelB}
{\langle \Pb : \dot \F -  \B \cdot \dot \p \rangle}={ \langle \Pb - \B \otimes \X \rangle : \langle \dot \F \rangle-  \langle \B \rangle \cdot \dot \p^\M }, 
\end{equation}
and of course, it reduces to the standard relation, $\langle \Pb : \dot \F \rangle= \langle \Pb \rangle : \langle \dot \F \rangle$, when the body forces and inertia terms vanish.

In the following, we proceed to prove that this relation holds exactly under a finite element discretization. The left hand side of the equation reads 
\begin{equation}
\begin{split}
\int_{\Omega_0} \left( P_{iJ} \dot \varphi_{i,J}^h -  B_i \dot \varphi_i^h \right) \,dV&= \int_{\Omega_0} P_{iJ} \sum_a \dot \varphi_{ai} N_{a,J} \, dV-\int_{\Omega_0}  B_i \sum_a \dot \varphi_{ai} N_{a} \, dV  \\
&=\sum_b \dot \varphi_{bi} \left[\int_{\Omega_0} \left(P_{iJ} N_{b,J}-B_i N_b \right) \, dV \right]  + \sum_c \dot \varphi_{ci} \left[\int_{\Omega_0} \left(P_{iJ} N_{c,J}-B_i N_c \right) \, dV \right]  \\
 &=  \sum_c \dot \varphi_{ci} \left[\int_{\Omega_0} \left(P_{iJ} N_{c,J}- B_i N_c \right) \, dV \right],
\end{split}
\end{equation}
where we have divided the full set of nodes into boundary nodes and interior nodes and have applied the equilibrium equation for the interior nodes, cf.~Eq.~\eqref{Eq:EqWeak_body}. Next, we make use of the boundary conditions given by Eq.~\eqref{BCC}, and obtain
\begin{equation}
\begin{split}
 \int_{\Omega_0} \left(P_{iJ} \dot \varphi_{i,J}^h \,dV- B_i \dot \varphi_i^h \right)\,dV  &=  \sum_c \dot \varphi^\M_{i} \left[\int_{\Omega_0} \left(P_{iJ} N_{c,J}-B_i N_c \right) \, dV \right]\\
& \phantom{=}+  \sum_c \dot F^\M_{iQ} X_{Qc} \left[\int_{\Omega_0} \left(P_{iJ} N_{c,J}- B_i N_c \right) \, dV \right]  \\ 
& \phantom{=}+ \sum_c  \dot {\tilde{\varphi}}_{ci} \left[\int_{\Omega_0} \left(P_{iJ} N_{c,J}- B_i N_c \right) \, dV \right], \\
\end{split}
\end{equation}
where the last term on the right hand side vanishes due to the periodicity of $\dot{\tilde{\varphi}}_{ci}$ and anti-periodicity of the term in brackets, cf.~Eq.\eqref{Eq:EqWeak_periodic}. Then, by using Eq.~\eqref{Eq:EqWeak_body} and extending the remaining sums to the full set of nodes \{$a$\}, one obtains
\begin{equation}
\begin{split}
&\int_{\Omega_0} \left( P_{iJ} \dot \varphi_{i,J}^h \,dV-B_i \dot \varphi_i^h \right)\,dV \\
&= \sum_a \dot \varphi^\M_{i} \left[\int_{\Omega_0} \left(P_{iJ} N_{a,J}-B_i N_a \right) \, dV \right]  + \sum_a \dot F^\M_{iQ} X_{Qa} \left[\int_{\Omega_0} \left(P_{iJ} N_{a,J}-B_i N_a \right) \, dV \right]  \\
& =  \dot \varphi^\M_{i} \bigg[\int_{\Omega_0} \Big(P_{iJ} \sum_a N_{a,J}- B_i \sum_a N_a \Big) \, dV \bigg]+  \dot F^\M_{iQ}  \bigg[\int_{\Omega_0} \Big(P_{iJ} \sum_a N_{a,J} X_{Qa}-B_i \sum_a N_a X_{Qa} \Big) \, dV \bigg].  \\
\end{split}
\end{equation}
Finally, the use of properties of the shape functions, cf.~Eqs.~\eqref{Eq:PartUnity} and \eqref{Eq:Linear}, delivers
\begin{equation}
\int_{\Omega_0} \left (P_{iJ} \dot \varphi_{i,J}^h -  B_i \dot \varphi_i^h \right) dV = \left[\int_{\Omega_0} \left(P_{iJ} -B_i X_J \right) \, dV \right] \dot F^\M_{iJ}-  \left[\int_{\Omega_0} B_i  \, dV \right] \dot \varphi^\M_{i}
\end{equation}
where we have further used the fact that Eq.~\eqref{Eq:PartUnity} implies $\sum_a N_{a,J}=0$.
This above result, together with the already derived expression $\F^\M=\langle \F \rangle$, completes the desired proof.

\subsection{Uniform traction boundary condition}

When inertial or body forces are considered, a modified averaging statement for the the first Piola-Kirchhoff stress tensor has been obtained in the smooth case \citep{molinari2001micromechanical,ricker2009multiscale,reina2011multiscale,pham2013transient,de2015rve}. It reads
\begin{equation} \label{Eq: Pave_general}
\mathbf{P}^\M = \langle \mathbf{P} - \mathbf{B} \otimes \mathbf{X}  \rangle.
\end{equation}
We show below that this equivalence holds exactly under a finite element discretization. 

\subsubsection*{Averaging statement for the first Piola-Kirchhoff stress tensor}
Using the properties of the shape functions, cf.~Eqs.~\eqref{Eq:Linear} and \eqref{DeltaP}, and the equilibrium equations for interior nodes $\{b\}$, cf.~Eq.~\eqref{Eq:EqWeak_body}, the right hand side of Eq.~\eqref{Eq: Pave_general} reads
\begin{equation}
\begin{split}
\int_{\Omega_0} \left(P_{iJ}-B_i X_J \right) dV &=\int_{\Omega_0}\left( P_{iQ}\delta_{QJ} -B_i X_J \right) dV = \int_{\Omega_0} \bigg( P_{iQ}\sum_a X_{aJ} N_{a,Q}-B_i \sum_a N_a X_{aJ}\bigg) dV\\
&= \sum_a X_{aJ} \int_{\Omega_0} \left( P_{iQ} N_{a,Q}-B_i  N_a \right) dV= \sum_c X_{cJ} \int_{\Omega_0} \left( P_{iQ} N_{c,Q}-B_i  N_c \right) dV.  \\
\end{split}
\end{equation}
 Then, by substituting the governing equations of boundary nodes $\{c\}$, cf.~Eq.~\eqref{Eq:EqWeak_traction}, we have
\begin{equation}
\begin{split}
\int_{\Omega_0} \left(P_{iJ}-B_i X_J \right) dV &=P_{iQ}^\M \sum_c X_{cJ} \int_{\partial \Omega_0} N_Q N_{c} \,dS.
\end{split}
\end{equation}
The shape functions associated to nodes $\{b\}$ are zero at $\partial \Omega_0$ and therefore the sum in the last term of the above equation may be extended over all nodes. Then, by the linear representation of the shape functions, cf.~Eq.~\eqref{Eq:Linear}, and application of the divergence theorem, the sought-after result is obtained
\begin{equation}
\begin{split}
\int_{\Omega_0} \left(P_{iJ}-B_i X_J \right) dV &= P_{iQ}^\M\sum_a X_{aJ}  \int_{ \partial \Omega_0}  N_Q N_{a} \, dS= P_{iQ}^\M  \int_{ \partial \Omega_0}  N_Q X_{J}  \, dS= P^\M_{iJ} |\Omega_0|.\\
\end{split}
\end{equation}

\subsubsection*{Hill-Mandel principle}
In this section, we proceed to derive Eq.~\eqref{Eq:HillMandelB} for the finite element solution of an RVE with uniform traction boundary condition. In this case, the finite element solution may be non-unique due to the undetermined value of the translations for the static problem, among other possible sources of non-uniqueness. Yet, these translations are important in the presence of body forces, as they contribute to the work these forces perform. It then results convenient to separate the nodal deformation mapping $\boldsymbol \varphi_a$ into the macroscopic translation, and the fluctuation around that value. Equivalently, $\boldsymbol \varphi^h=\boldsymbol \varphi^\M+\hat {\boldsymbol \varphi}^h$, which satisfies $\nabla \boldsymbol \varphi^h= \nabla \hat{\boldsymbol  \varphi}^h$. Then, using the weak form of the equilibrium equations associated to nodes  $\{b\}$, cf.~Eq.~\eqref{Eq:EqWeak_body}, it follows that
\begin{equation}
\begin{split}
\int_{\Omega_0} \left( P_{iJ} \dot \varphi_{i,J}^h - B_i \dot \varphi_{i}^h \right) dV &=\int_{\Omega_0}P_{iJ}\sum_a \dot {\hat \varphi}_{ai}N_{a,J} \, dV - \int_{\Omega_0} B_i \bigg( \sum_a \dot {\hat \varphi}_{ai} N_a +\dot \varphi_i^\M \bigg)dV \\
&=\sum_a \dot {\hat \varphi}_{ai}\int_{\Omega_0} \left( P_{iJ}N_{a,J} \, dV-B_i N_a \right) dV-\int_{\Omega_0} B_i  \dot {\varphi}_i^\M \, dV \\
&=\sum_c \dot {\hat \varphi}_{ci}\int_{\Omega_0} \left( P_{iJ}N_{c,J} \, dV-B_i N_c \right) dV - \dot \varphi_i^\M\int_{\Omega_0} B_i \, dV. \\
\end{split}
\end{equation}
Then, by substituting the governing equations of boundary nodes $\{c\}$, cf.~Eq.~\eqref{Eq:EqWeak_traction}, we obtain 
\begin{equation}
\begin{split}
\int_{\Omega_0} \left( P_{iJ} \dot \varphi_{i,J}^h - B_i \dot \varphi_{i}^h \right) dV &=\sum_c \dot {\hat \varphi}_{ci}\int_{\partial \Omega_0} P_{iQ}^\M N_Q N_{c}dS- \dot \varphi_i^\M\int_{\Omega_0} B_i  \, dV\\
&=P_{iQ}^\M \sum_c \dot {\hat \varphi}_{ci}\int_{\partial \Omega_0}  N_Q N_{c}\,dS- \dot \varphi_i^\M\int_{\Omega_0} B_i  \, dV.\\
\end{split}
\end{equation}
The shape functions associated to nodes $\{b\}$ are zero at $\partial \Omega_0$, and, therefore
\begin{equation}
\begin{split}
\int_{\Omega_0} \left( P_{iJ} \dot \varphi_{i,J}^h - B_i \dot \varphi_{i}^h \right) dV &=P_{iQ}^\M \sum_a \dot {\hat \varphi}_{ai}\int_{\partial \Omega_0}  N_Q N_{a}\,dS- \dot \varphi_i^\M\int_{\Omega_0} B_i \, dV. \\
\end{split}
\end{equation}
Finally, application of divergence theorem gives,
\begin{equation}
\begin{split}
\int_{\Omega_0} \left( P_{iJ} \dot \varphi_{i,J}^h - B_i \dot \varphi_{i}^h \right) dV&=P_{iJ}^\M \int_{\partial \Omega_0} \hat \varphi^h_i  N_J\,dS - \dot \varphi_i^\M\int_{\Omega_0} B_i \, dV \\
&= P_{iJ}^\M \int_{\Omega_0} \dot \varphi_{i,J}^h \, dV- \dot \varphi_i^\M\int_{\Omega_0} B_i  \,dV.  \\
\end{split}
\end{equation}
Substituting the previously obtained relation $\Pb^\M= \langle \Pb-\B\otimes \X \rangle$, we obtain the desired relation
\begin{equation}
\langle \Pb : \dot \F -  \B \cdot \dot \p \rangle= \langle \Pb - \B \otimes \X \rangle : \langle \dot \F \rangle-  \langle \B \rangle \cdot \dot \p^\M.
\end{equation}

\section{Numerical example}
\begin{figure} [H]
\begin{center}  
\subfigure[RVE model] { \label{fig:a}    
\includegraphics[width=0.381\columnwidth]{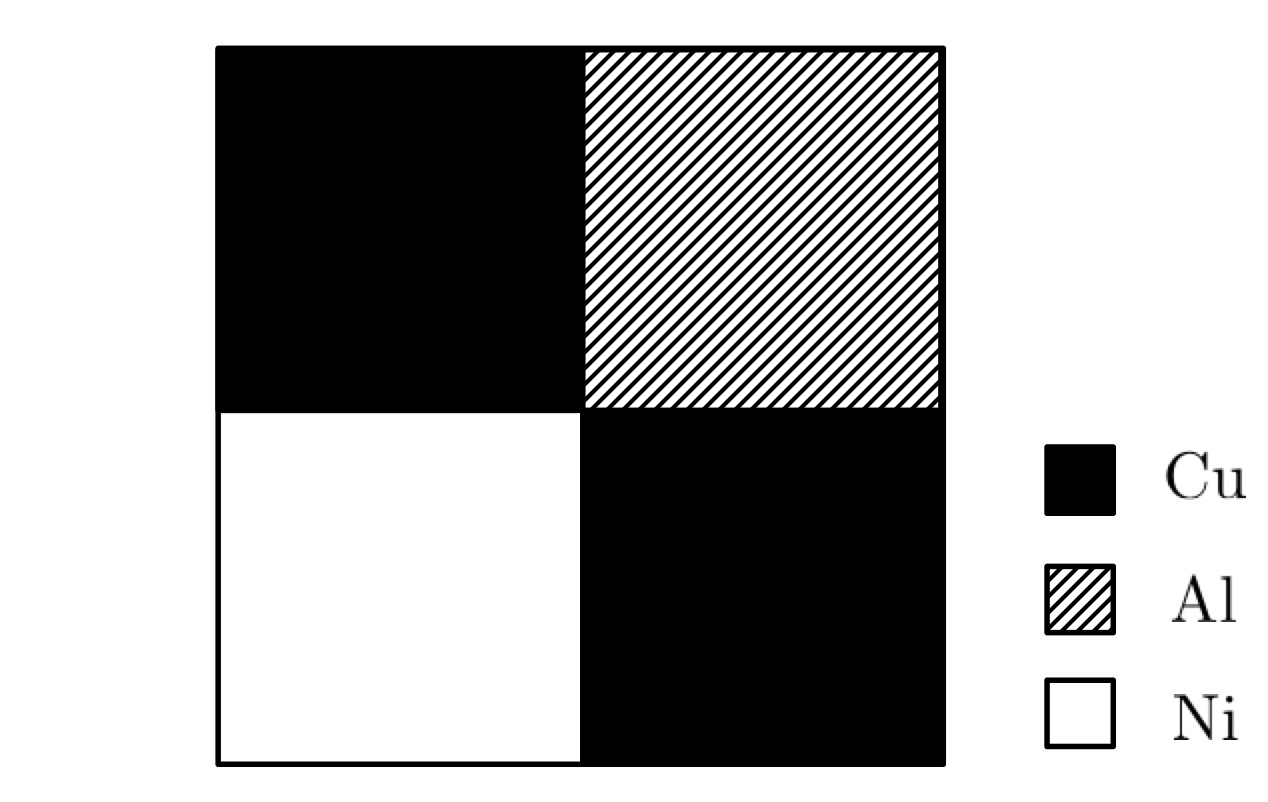} 
}    \qquad
\subfigure[Mesh distribution] { \label{fig:b}    
\includegraphics[width=0.238\columnwidth]{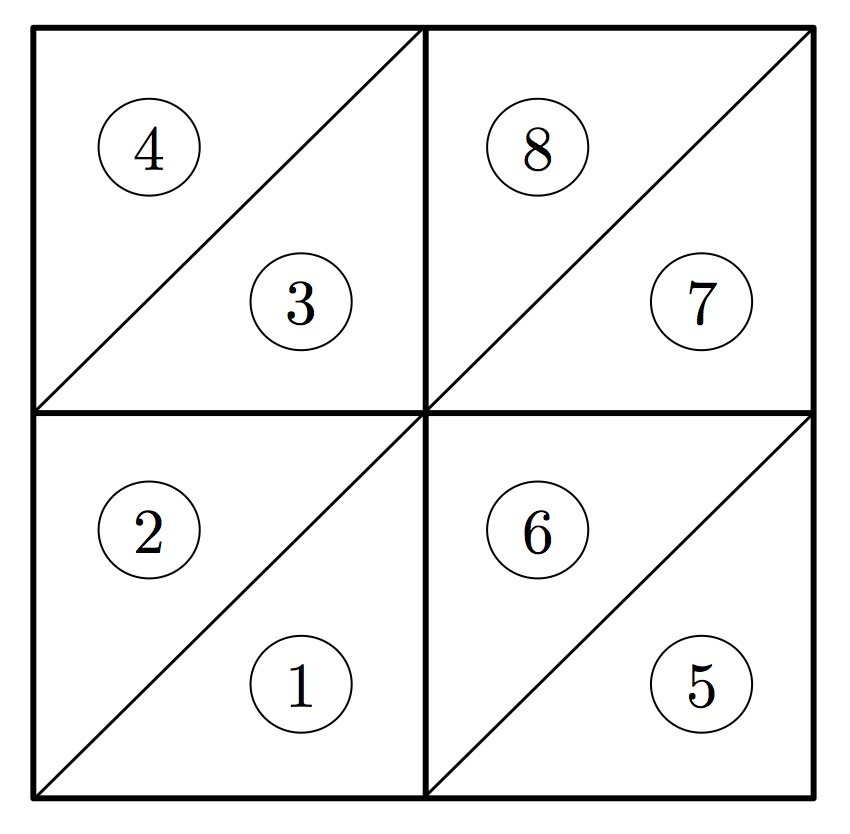}    
}    
\caption{Geometry of RVE and undeformed mesh distribution.} \label{Sche}    
\end{center}
\end{figure}

In this section, the volume-averaged relations are verified numerically over an RVE for the three types of boundary conditions considered (affine, periodic, and traction boundary conditions), and two modes of deformation (simple extension and simple shear). 
Figure \ref{Sche}(a) shows the domain of the RVE, which is a unit square composed of three materials: Al, Cu and Ni. These materials are arranged with no geometric symmetry, providing a general and irregular form of RVE. Furthermore, a coarse mesh is intentionally employed to highlight the validity of the results, even when far away from the smooth exact solution. In particular, the microscopic domain is divided into 8 identical linear triangular elements, as shown in Figure \ref{Sche}(b), and linear shape functions are used for the finite element discretization. Simulations are performed using COMSOL multiphysics software, where hyperelastic constitutive relations (neo-Hookean) are used for the three materials, whose properties are listed in Table \ref{Materials}. The details of the simulations are provided below along the text. 

\begin{table}[H] 
\footnotesize
\begin{center}  
\caption{Materials properties for Al, Cu, and Ni from COMSOL material library. } 
\label{Materials}
\begin{tabular}{l*{7}{c}r} \\[-1.70ex]
\hline
                 & Al  & Cu  & Ni \\
\hline
Density $\rho$ (kg/$\text{m}^3$)            & 2700                                        & 8960                & 8900             \\[-1.70ex]
Lam\'e constant $\lambda$ (GPa)          & 60.49                                       & 95.15                & 136.4              \\[-1.70ex]
Lam\'e constant $\mu$ (GPa)                & 25.93                                        & 44.78                & 80.15               \\
\hline
\end{tabular}
\end{center}
\end{table}\label{1}

\subsection{Affine and periodic displacement boundary condition}
\begin{figure} [H]
\begin{center}  
\subfigure[Affine displacement b.c.] { \label{fig:a}    
\includegraphics[width=0.3\columnwidth]{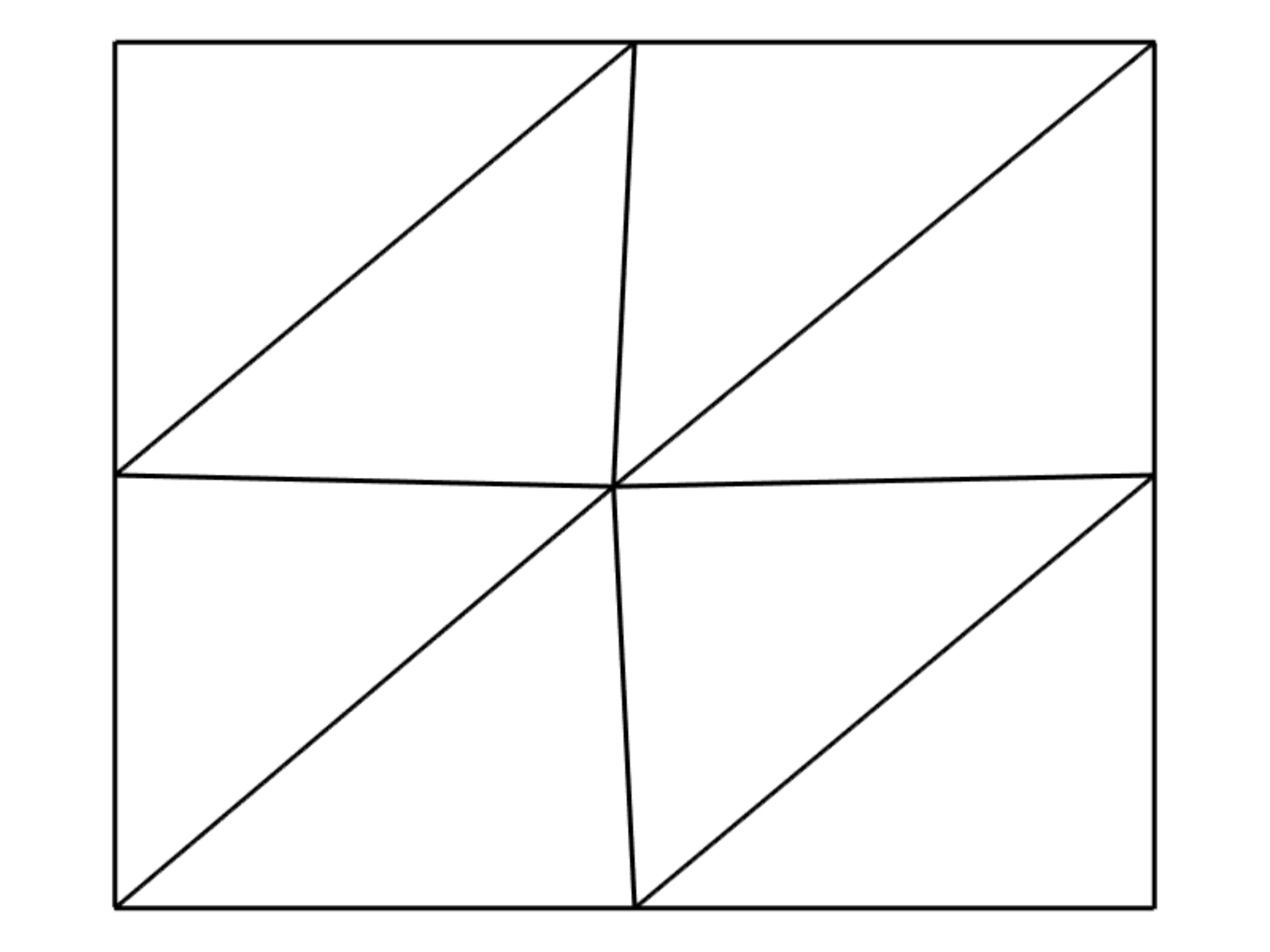} 
}    \qquad
\subfigure[Periodic b.c.] { \label{fig:b}    
\includegraphics[width=0.3\columnwidth]{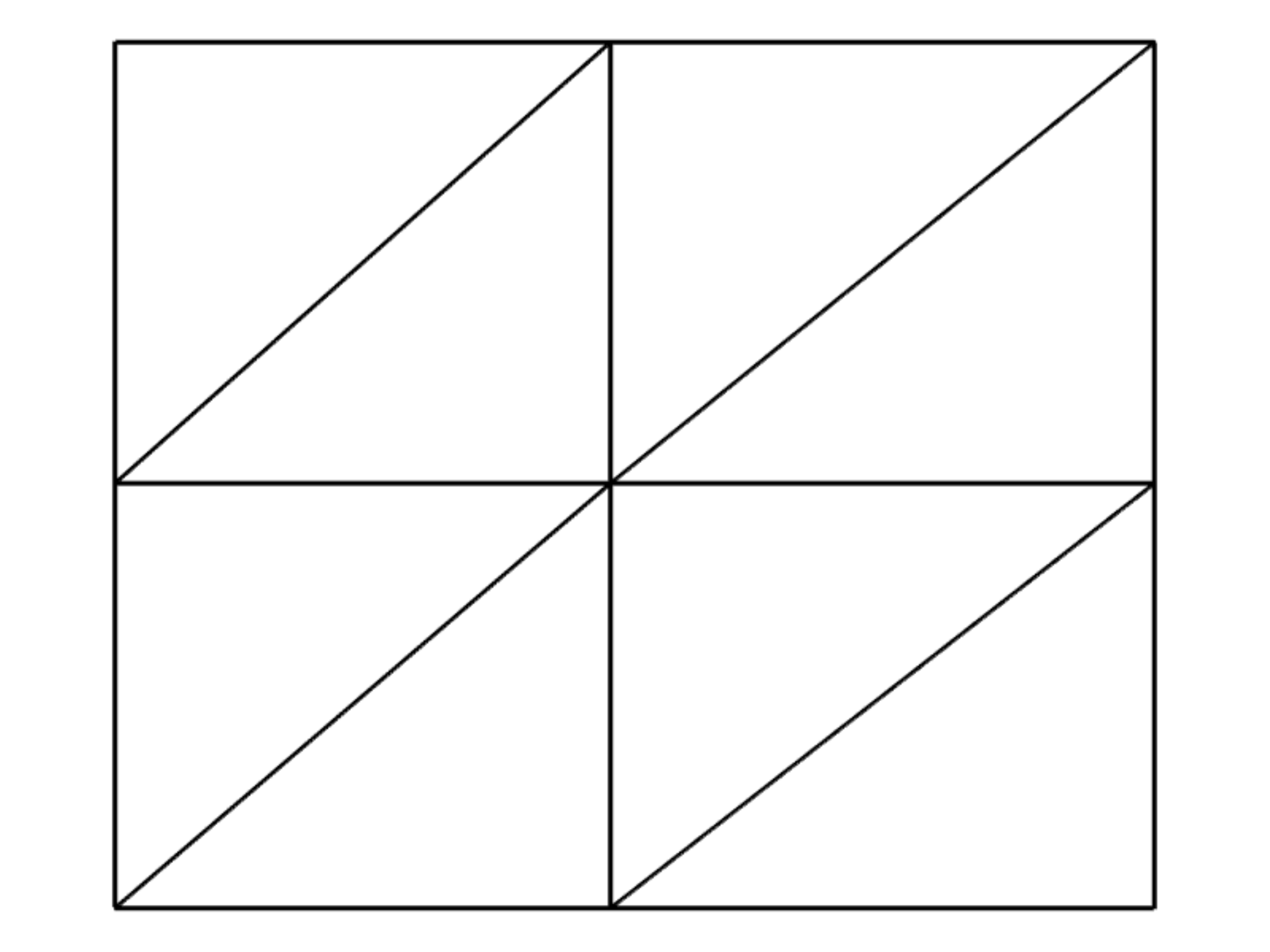}    
}    
\caption{Deformed meshes for simple extension mode under affine and periodic boundary conditions (b.c.).} \label{ExD}    
\end{center}
\end{figure}

We begin by considering the plane-strain extension of the the RVE previously described according to the macroscopic deformation gradient $\F^\M=[F_{11}^\M, F_{12}^\M; F_{21}^\M, F_{22}^\M]=[1.2, 0; 0, 1]$. Computationally, 
this macroscopic deformation is applied in 20 uniform steps, and at each of these steps, the stress and deformation gradient of each element is computed. Tables \ref{ExLin} and \ref{ExPeri} show the values of each component of $\F$ and $\Pb$ for each element in the final configuration, represented in Figures \ref{ExD}(a) and (b), for the affine and the periodic boundary conditions, respectively.

As the results of Tables \ref{ExLin}(b) and \ref{ExPeri}(b) indicate, the average relation $\F^{\M}=\langle \F \rangle$ is satisfied exactly, up to machine precision, for both affine and periodic boundary conditions. That is, 
\begin{equation}
\frac{1}{|\Omega_0|}\int_{\Omega_0}\F \,dV=
\begin{bmatrix}
    \frac{1}{8}\sum_{e=1}^8 (F_{11})_e & \frac{1}{8}\sum_{e=1}^8 (F_{12})_e  \\[1.70ex]
    \frac{1}{8}\sum_{e=1}^8 (F_{21})_e & \frac{1}{8}\sum_{e=1}^8 (F_{22})_e 
  \end{bmatrix}  
  =\begin{bmatrix}
    \,1.2 &  0\,\\
    \,0 & 1\,
  \end{bmatrix} ,
\end{equation}
where the sum is performed over the elements of the triangular mesh. A similar calculation can be carried out at each time step for the deformation gradient and the stress tensor, allowing us to verify the Hill-Mandel principle. For the linear displacement boundary conditions, the volumetric average of the strain energy stored during the deformation process is
\begin{equation}
\int_{t_0}^{t}\langle \mathbf{P}: \dot\F \rangle dt=\int_{t_0}^{t} \bigg( \frac{1}{|\Omega_0|} \int _{\Omega_0} \mathbf{P} :\dot\F dV \bigg) dt=3.758512926\E9\,\text{Pa},
\end{equation}
while the energy evaluated from the macroscopic (averaged) quantities leads to
\begin{equation}
 \int_{t_0}^{t}\langle \mathbf{P} \rangle: \langle \dot \F \rangle \,dt= \int_{t_0}^{t}\langle \mathbf{P} \rangle: \dot \F^\M \,dt=\int_{t_0}^{t} \bigg( \frac{1}{|\Omega_0|}\int _{\Omega_0}\mathbf{P}: \dot\F^\M dV \bigg) dt=3.758512926\E9\,\text{Pa},
\end{equation}
which has, as expected, an identical value. Similarly, for the periodic boundary condition, the Hill-Mandel principle is verified exactly, with $\langle \mathbf{P}: \dot\F \rangle =  \langle \mathbf{P} \rangle : \langle \dot \F \rangle=3.638415107\E9$\,\text{Pa}.

\begin{figure} [H]
\begin{center}  
\subfigure[Affine displacement b.c.] { \label{fig:a}    
\includegraphics[width=0.3\columnwidth]{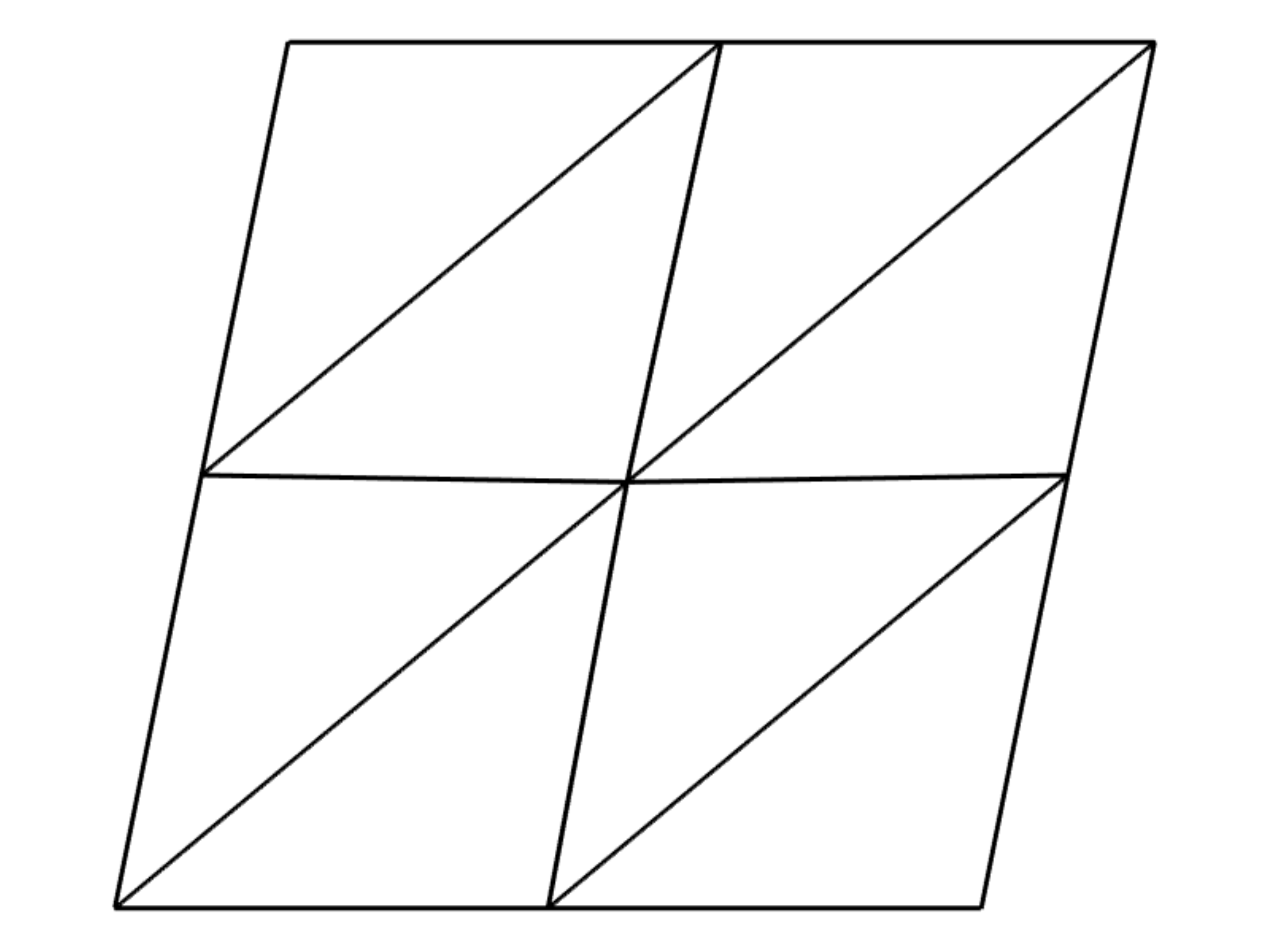} 
}    \qquad
\subfigure[Periodic b.c.] { \label{fig:b}    
\includegraphics[width=0.3\columnwidth]{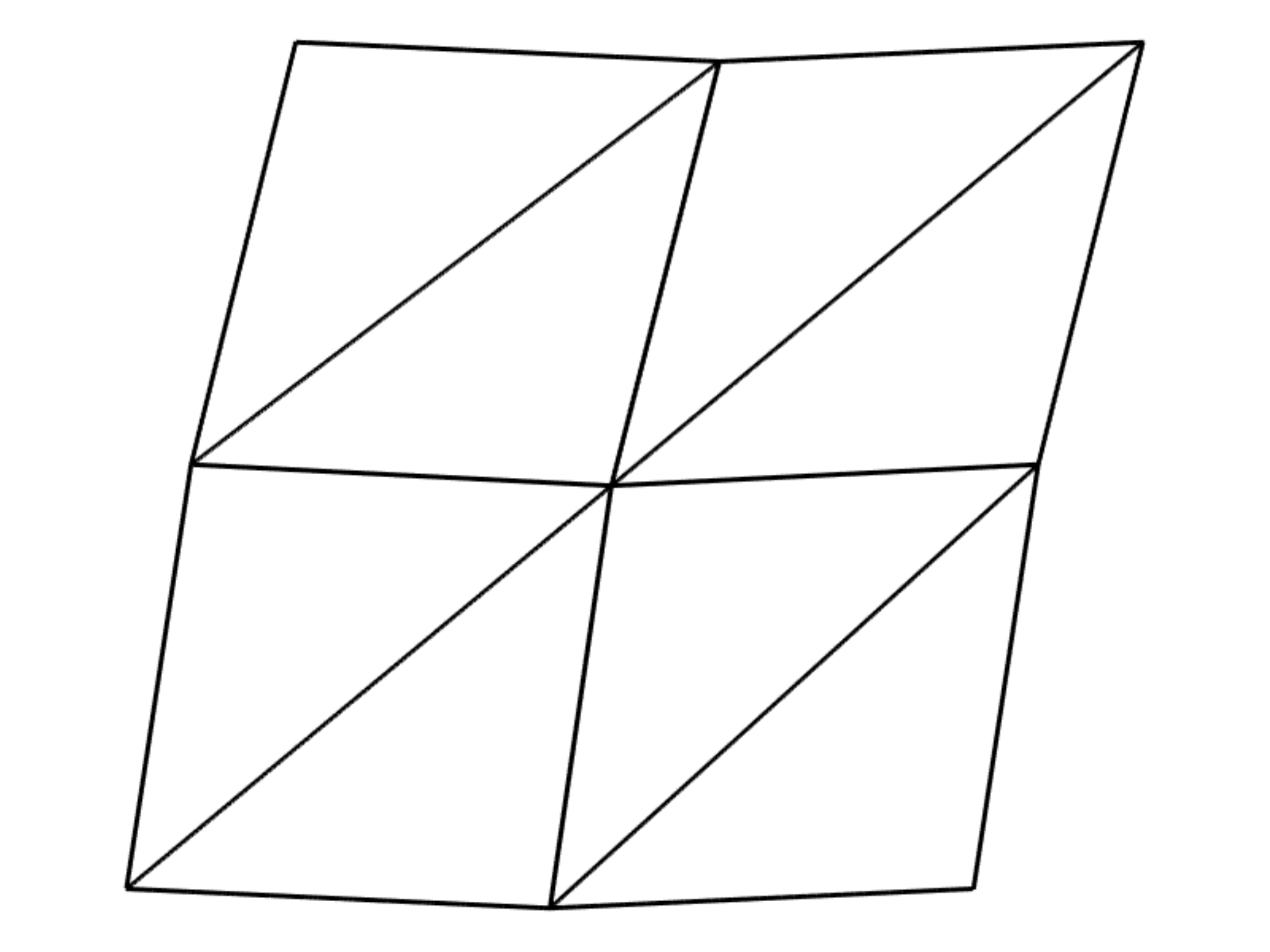}    
}    
\caption{ Deformed meshes for simple shear mode under affine and periodic boundary conditions (b.c.).} \label{Sh}    
\end{center}
\end{figure}

A similar verification analysis is performed for a simple shear mode according to 
the macroscopic deformation gradient $\F^\M=[F_{11}^\M, F_{12}^\M; F_{21}^\M, F_{22}^\M]=[1, 0.2; 0, 1]$. The corresponding deformed meshes of RVE are shown in Figure \eqref{Sh} for the affine and periodic boundary conditions, respectively, and the components of the stress and deformation gradient tensor for the final configuration are recorded in Tables \ref{ShLin} and \ref{ShPeri} for each element of the mesh. As indicated in the tables, the average deformation gradient coincides for both types of boundary conditions with the imposed macroscopic deformation, verifying the averaging relation $\F^{\M}=\langle \F \rangle$.

Similarly, Hill-Mandel principle holds exactly under the finite element discretization. For linear displacement boundary condition, $\int_{t_0}^{t}\langle \mathbf{P}: \dot\F \rangle \,dt= \int_{t_0}^{t} \langle \mathbf{P} \rangle : \langle \dot \F \rangle \, dt=9.569752719\E8\,\text{Pa}$, and for periodic boundary conditions, $\int_{t_0}^{t} \langle \mathbf{P}: \dot\F \rangle\, dt=\int_{t_0}^{t} \langle \mathbf{P} \rangle : \langle \dot \F \rangle \,dt=8.395522388\E8\,\text{Pa}$.


\subsection{Uniform traction boundary conditions}
\begin{figure} [H]
\begin{center}  
\subfigure[Simple extension mode] { \label{fig:a}    
\includegraphics[width=0.3\columnwidth]{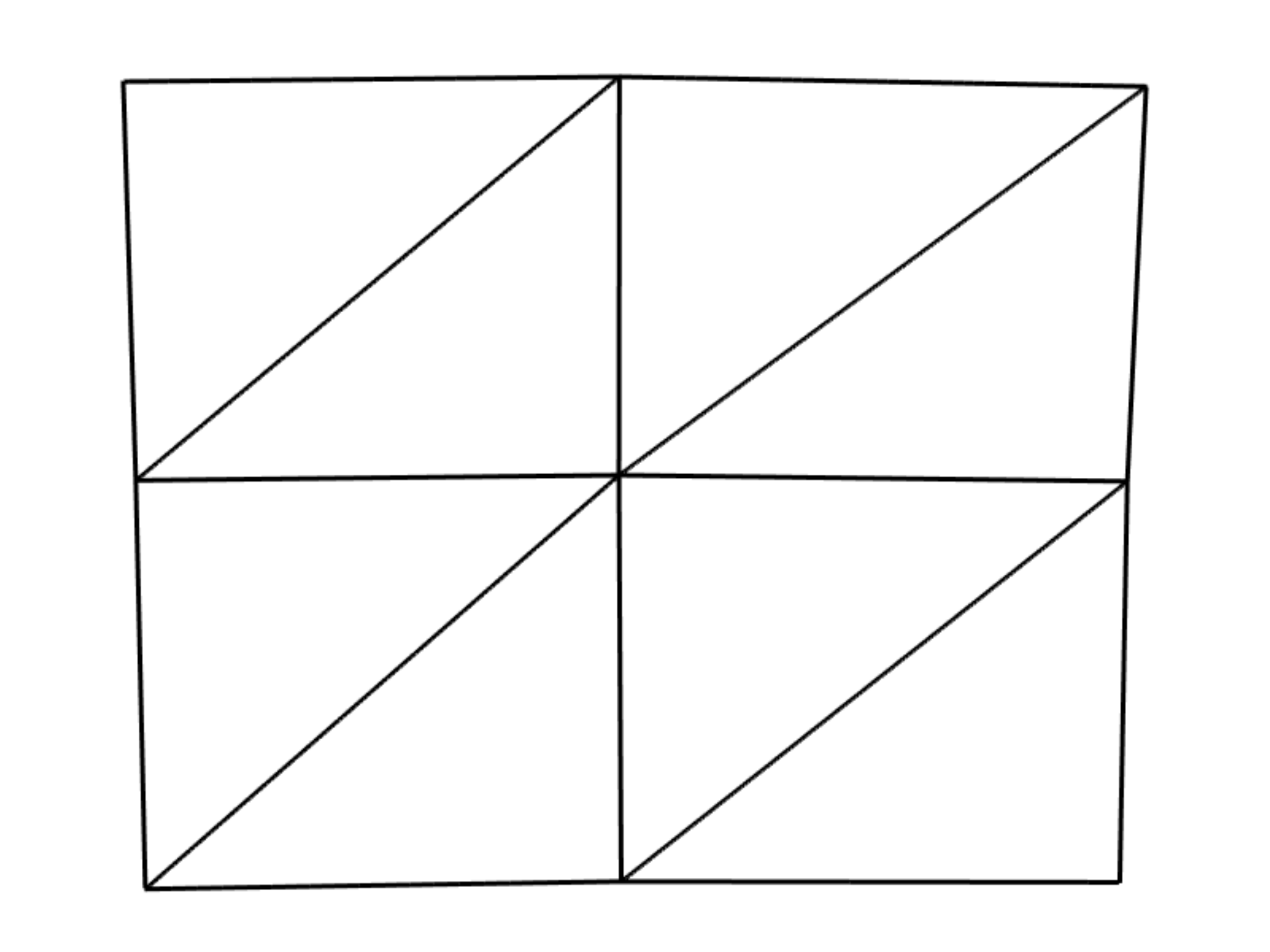} 
}    \qquad
\subfigure[Simple shear mode] { \label{fig:b}    
\includegraphics[width=0.3\columnwidth]{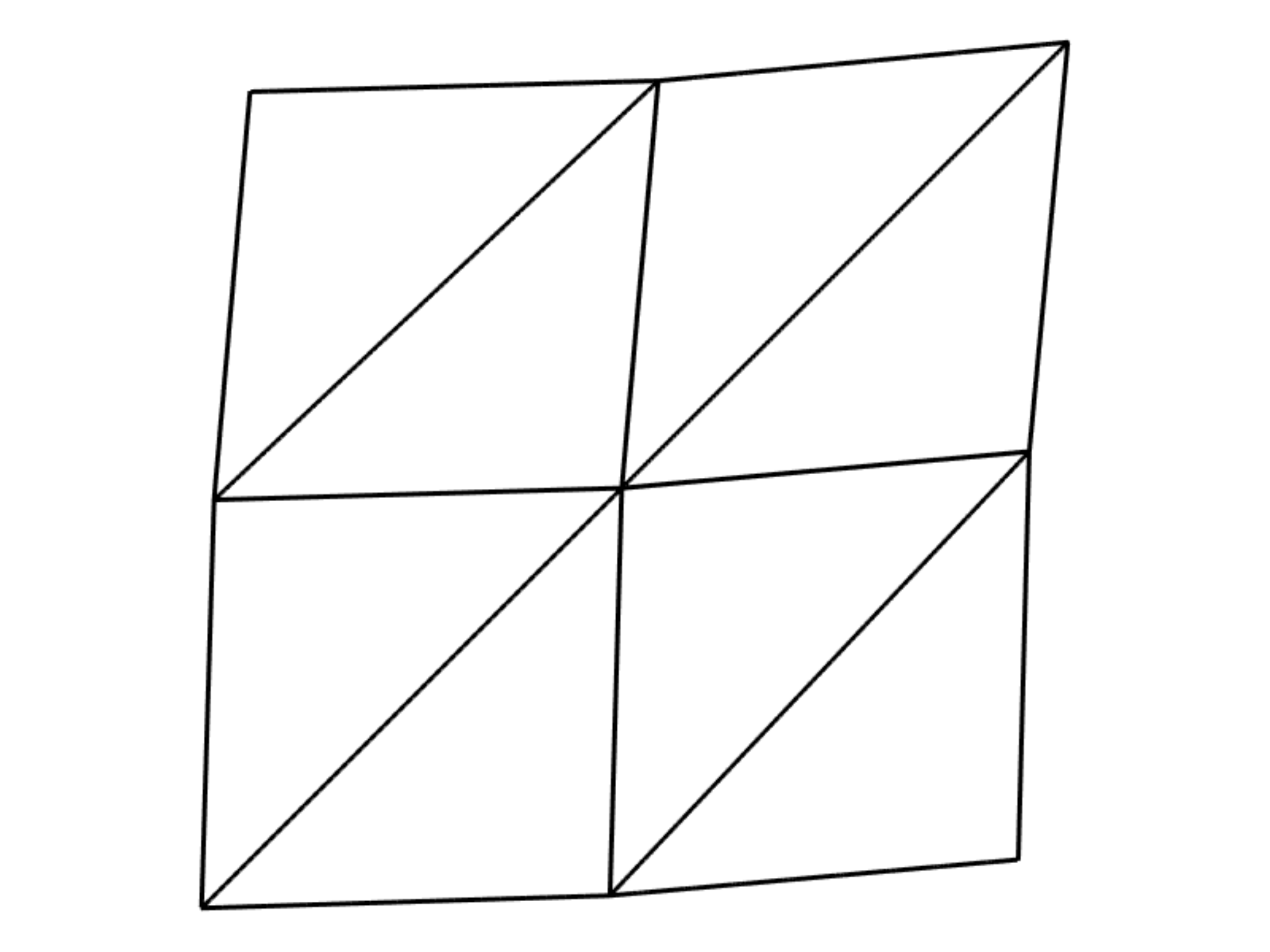}    
}    
\caption{Deformed meshes for simple extension mode and simple shear mode under uniform traction boundary condition.} \label{Trac}    
\end{center}
\end{figure}

Next, we examine the averaging relations when the RVE is subjected to uniform traction boundary conditions. In particular, the following examples consider an applied macroscopic traction of $\mathbf P^\M=[P_{11}^\M, P_{12}^\M; P_{21}^\M, P_{22}^\M]$=[1, 0; 0, 0] GPa for a simple extension mode and $\mathbf P^\M$=[$P_{11}^\M$, $P_{12}^\M$; $P_{21}^\M$, $P_{22}^\M$]=[0, 1; 1, 0] GPa for a simple shear mode. The loading step size is set to 0.1GPa when the non-zero stress component is smaller than 0.4 GPa, and to 0.01GPa in the following range. Figure \ref{Trac}(a) and (b) shows the deformed meshes of RVE for both modes of deformation. 

The components of the stress tensor and the deformation tensor at the last step of deformation for all the elements have been listed in Tables \ref{ExTrac} and \ref{ShTrac} for simple extension mode and simple shear mode, respectively.  The results indicate that the volume average of the stress tensor is exactly equal to $\Pb^{\M}$, up to machine precision, verifying the discrete stress averaging result derived in Section \ref{Sec:UniformTraction_noB}.
%
%
Similarly, the Hill-Mandel principle is verified numerically. For simple extension mode, the micro- and macro-work coincide and are equal to $\int_{t_0}^{t} \langle \mathbf{P}: \dot\F \rangle\, dt=\int_{t_0}^{t} \langle \mathbf{P} \rangle : \langle \dot \F \rangle \,dt=3.726695734\E6\,\text{Pa}$, whereas for the simple shear mode, $\int_{t_0}^{t}\langle \mathbf{P}: \dot\F \rangle \,dt=\int_{t_0}^{t} \langle \mathbf{P} \rangle : \langle \dot \F \rangle \,dt=1.192942091\E7\,\text{Pa}$.

\subsection{Discussion}

The numerical examples above confirm the validity of the stress and strain averaging relations as well as Hill-Mandel principle for a finite element discretization regardless of the size of the RVE or the mesh size. The energetic consistency relation, i.e.~$\langle \mathbf{P}: \dot\F \rangle = \langle \mathbf{P} \rangle : \langle \dot \F \rangle$, is demonstrated for stresses and strains that belong to the same boundary value problem, as this is often its primary use. Yet, this relation only requires a compatible deformation gradient and a stress tensor in equilibrium that share the same finite element discretization. However, they are not required to be related to each other, as shown in the analytical derivations.

\section{\large Conclusion}
In summary, the averaging relations needed to establish the consistency between the two levels of description in computational homogenization methods (FE$^2$) are shown to be exact under standard finite element discretizations. These are the stress and strain averaging relations as well as Hill-Mandel principle of energetic consistency. The proofs are performed in the finite kinematic setting, for generality, and are shown to hold for the three standard type of boundary conditions (affine displacement, uniform stress and periodic boundary conditions) for static or dynamic loading.
The analytical proofs are further supported by numerical examples of a highly asymmetric RVE with a coarse finite element discretization. This work provides a solid foundation to the applications of Hill's averaging theorems in RVE-based multiscale finite element methods with conventional Dirichlet, Neumann or periodic boundary conditions.


\section*{\large Acknowledgements}
C. Reina acknowledges discussions with Prof. Michael Ortiz.

\section*{\large Appendix A}

\begin{subtables}\label{ExLin}
\begin{table}[H]\footnotesize
\begin{center}  
    \caption{Components of stress tensor for simple extension mode under linear displacement boundary condition.} 
    \begin{tabular}{l*{7}{c}r} \\[-1.70ex]
\hline
Number             & $P_{11}$(Pa)  & $P_{12}$(Pa)  & $P_{21}$(Pa)  & $P_{22}$(Pa)   \\
\hline
1             & 5.540445662129311E10    & -3.697775921110617E9      & -3.697775921110617E9       & 1.872085738015028E10  \\[-1.70ex]
2             & 4.564975169177103E10    & -2.311265268571417E9     & -2.311265268571417E9        & 2.098423482999226E10    \\[-1.70ex]
3             & 3.116295439474504E10    & 7.745539252061372E8       & 7.745539252061372E8        & 1.996623282454363E10   \\[-1.70ex]
4             & 3.694029850600000E10    & 0                                          & 0                                           & 1.902985074600000E10   \\[-1.70ex]
5             & 3.694029850600000E10    & 0                                          & 0                                           & 1.902985074600000E10  \\[-1.70ex]
6             & 4.271764261725496E10    & -7.745539252061372E8    & -7.745539252061372E8        & 1.809346866745638E10 \\[-1.70ex]
7             & 2.765210896757578E10    & 7.475944449735475E8      & 7.475944449735475E8        & 1.488959713525095E10   \\[-1.70ex]
8             & 2.421352284198325E10    & 1.196070730162412E9      & 1.196070730162412E9        & 1.533834135993035E10   \\
\hline
\end{tabular}
\end{center}
\end{table}
\begin{table}[H]\footnotesize
\begin{center}  
\caption{Components of deformation gradient for simple extension mode under linear displacement boundary condition.} 
\begin{tabular}{l*{7}{c}r} \\[-1.70ex]
\hline
Number             & $F_{11}$  & $F_{12}$  & $F_{21}$  & $F_{22}$   \\
\hline
1             & 1.2                                    & -0.04613415672758626        & 0                                                 & 0.9711642142698373  \\[-1.70ex]
2             & 1.153865843272414        & 0                                            & -0.028835785730162634           & 1                                   \\[-1.70ex]
3             & 1.153865843272414        & 0.04613415672758626         & -0.028835785730162634           & 1.028835785730163   \\[-1.70ex]
4             & 1.2                                    & 0                                            & 0                                                 & 1                                  \\[-1.70ex]
5             & 1.2                                    & 0                                            & 0                                                  & 1                                   \\[-1.70ex]
6             & 1.246134156727586        & -0.04613415672758626        & 0.028835785730162634             & 0.9711642142698373 \\[-1.70ex]
7             & 1.246134156727586        & 0                                            & 0.028835785730162634             & 1                                   \\[-1.70ex]
8             & 1.2                                    & 0.04613415672758626         & 0                                                  & 1.028835785730163    \\
\hline
Average  & 1.200000000000000        & 0                                            & 0                                                  & 1 \\
\hline
\end{tabular}
\end{center}
\label{table:1}
\end{table}
\end{subtables}

\begin{subtables}\label{ExPeri}
\begin{table}[H]\footnotesize
\begin{center}  
\caption{Components of stress tensor for simple extension mode under periodic boundary condition.} 
\begin{tabular}{l*{7}{c}r} \\[-1.70ex]
\hline
Number             & $P_{11}$(Pa)  & $P_{12}$(Pa)  & $P_{21}$(Pa)  & $P_{22}$(Pa)   \\
\hline
1             & 4.272877655616882E10    & 6.532528423931364E-7       & 6.532528423931364E-7       & 1.517259313986339E10  \\[-1.70ex]
2             & 4.272877655616881E10    & 2.780854427286705E-7       & 2.780854427286705E-7       & 1.517259313986339E10   \\[-1.70ex]
3             & 3.003952559993056E10    & 1.398135338753770E-6       & 1.398135338753770E-6        & 1.806628499750840E10   \\[-1.70ex]
4             & 3.003952559993057E10    & 1.297020385082037E-6       & 1.297020385082037E-6        & 1.806628499750839E10   \\[-1.70ex]
5             & 4.384107141206944E10    & -5.423341730090870E-8     & -5.423341730090870E-8        & 1.999341649449161E10  \\[-1.70ex]
6             & 4.384107141206944E10    & -1.242786967781129E-6     & -1.242786967781129E-6        & 1.999341649449161E10 \\[-1.70ex]
7             & 2.892723074402995E10    & -8.994862469441367E-7     & -8.994862469441367E-7        & 1.709972463684661E10   \\[-1.70ex]
8             & 2.892723074402994E10    & -3.140181817261574E-8     & -3.140181817261574E-8        & 1.709972463684661E10   \\
\hline
\end{tabular}
\end{center}
\label{table:1}
\end{table}
\begin{table}[H]\footnotesize
\begin{center}  
\caption{Components of deformation gradient for simple extension mode under periodic boundary condition.} 
\begin{tabular}{l*{7}{c}r} \\[-1.70ex]
\hline
Number             & $F_{11}$  & $F_{12}$  & $F_{21}$  & $F_{22}$   \\
\hline
1             & 1.152799641638063    & 0                                        & 0        & 0.9809015450978975  \\[-1.70ex]
2             & 1.152799641638063   & 0                                         & 0        & 0.9809015450978975                                     \\[-1.70ex]
3            & 1.152799641638063    & 0                                         & 0        & 1.019098454902103  \\[-1.70ex]
4            & 1.152799641638063     & 0                                        & 0        & 1.019098454902103   \\[-1.70ex]
5             & 1.247200358361937     & 0                                      & 0         & 0.9809015450978975  \\[-1.70ex]
6             & 1.247200358361937     & 0                                      & 0         & 0.9809015450978975\\[-1.70ex]
7            & 1.247200358361937      & 0                                      & 0         & 1.019098454902103   \\[-1.70ex]
8            & 1.247200358361937       & 0                                     & 0         & 1.019098454902103   \\
\hline
Average & 1.200000000000000      & 0                                     & 0          & 1 \\
\hline
\end{tabular}
\end{center}
\label{table:1}
\end{table}
\end{subtables}

\begin{subtables}\label{ShLin}
\begin{table}[H]\footnotesize
\begin{center}  
\caption{Components of stress tensor for simple shear mode under linear displacement boundary condition.} 
\begin{tabular}{l*{7}{c}r} \\[-1.70ex]
\hline
Number             & $P_{11}$(Pa)  & $P_{12}$(Pa)  & $P_{21}$(Pa)  & $P_{22}$(Pa)   \\
\hline
1             & -2.130595379666259E9    & 1.477835058908113E10        & 1.477835058908113E10       & -4.634962905504009E9  \\[-1.70ex]
2             & -4.634962905504009E9    & 1.477835058908113E10        & 1.477835058908113E10       & -2.130595379666260E9    \\[-1.70ex]
3             & -1.399028340492011E9    & 8.955223880000000E9          & 8.955223880000000E9           & 1.399028340492011E9   \\[-1.70ex]
4             &  0                                       & 8.955223880000000E9          & 8.955223880000000E9           & 0   \\[-1.70ex]
5             &  0                                       & 8.955223880000000E9          & 8.955223880000000E9           & 0  \\[-1.70ex]
6             & 1.399028340492011E9    & 8.955223880000000E9           & 8.955223880000000E9           & -1.399028340492011E9 \\[-1.70ex]
7             & 1.755118475952594E9    & 5.590212526640496E9           & 5.590212526640496E9         & 9.450637946716006E8   \\[-1.70ex]
8             & 9.450637946716006E8    & 5.590212526640497E9           & 5.590212526640497E9         & 1.755118475952594E9   \\
\hline
\end{tabular}
\end{center}
\label{table:1}
\end{table}
\begin{table}[H]\footnotesize
\begin{center}  
\caption{Components of deformation gradient for simple shear mode under linear displacement boundary condition.} 
\begin{tabular}{l*{7}{c}r} \\[-1.70ex]
\hline
Number             & $F_{11}$  & $F_{12}$  & $F_{21}$  & $F_{22}$   \\
\hline
1             & 1                                         & 0.1843775168634644           & 0                                              & 0.9843775168634644  \\[-1.70ex]
2             & 0.9843775168634644        & 0.2                                        & -0.01562248313653562           & 1                                       \\[-1.70ex]
3            & 0.9843775168634644         & 0.2156224831365357          & -0.01562248313653562           & 1.015622483136536   \\[-1.70ex]
4            & 1                                          & 0.2                                         & 0                                              & 1                                         \\[-1.70ex]
5             & 1                                         & 0.2                                         & 0                                              & 1                                          \\[-1.70ex]
6             & 1.015622483136536          & 0.1843775168634644           & 0.01562248313653562            & 0.9843775168634644     \\[-1.70ex]
7            & 1.015622483136536           & 0.2                                         & 0.01562248313653562            & 1                                          \\[-1.70ex]
8            & 1                                          & 0.2156224831365357           & 0                                               & 1.015622483136536      \\
\hline
Average  & 1                                        & 0.2000000000000000            & 0                                               & 1 \\
\hline
\end{tabular}
\end{center}
\label{table:1}
\end{table}
\end{subtables}

\begin{subtables}\label{ShPeri}
\begin{table}[H]\footnotesize
\begin{center}  
\caption{Components of stress tensor for simple shear mode under periodic boundary condition.} 
\begin{tabular}{l*{7}{c}r} \\[-1.70ex]
\hline
Number             & $P_{11}$(Pa)  & $P_{12}$(Pa)  & $P_{21}$(Pa)  & $P_{22}$(Pa)   \\
\hline
1              & 1.892653704593794E-6    & 7.835820896554924E9     & 7.835820896554924E9     & 4.117337246423158E-6  \\[-1.70ex]
2              & 7.064054015297273E-7    & 7.835820896554925E9     & 7.835820896554925E9     & 1.536736099019807E-6    \\[-1.70ex]
3             & -6.602305766684191E-7    & 8.955223880000004E9     & 8.955223880000004E9    & -1.281624060558984E-6   \\[-1.70ex]
4             &  4.633653368252543E-6    & 8.955223880000004E9     & 8.955223880000004E9     & 1.684227316024411E-6   \\[-1.70ex]
5             &  4.928428739833904E-7    & 8.955223880000000E9     & 8.955223880000000E9     & 9.566949906492653E-7  \\[-1.70ex]
6              & 3.883709277785474E-8    & 8.955223880000000E9     & 8.955223880000000E9     & 1.903017544449548E-6 \\[-1.70ex]
7             & -1.092893533105891E-6    & 7.835820896554928E9     & 7.835820896554928E9    & -1.001675136507230E-6   \\[-1.70ex]
8             & -3.537979237633326E-6    & 7.835820896554928E9     & 7.835820896554928E9    & -2.458595741300362E-6   \\
\hline
\end{tabular}
\end{center}
\label{table:1}
\end{table}

\begin{table}[H]\footnotesize
\begin{center}  
\caption{Components of deformation gradient for simple shear mode under periodic boundary condition.} 
\begin{tabular}{l*{7}{c}r} \\[-1.70ex]
\hline
Number             & $F_{11}$  & $F_{12}$  & $F_{21}$  & $F_{22}$   \\
\hline
1            & 1                                          & 0.1488805970187594                & -0.05111940298124053             & 1 \\[-1.70ex]
2            & 1                                          & 0.1488805970187594                & -0.05111940298124052             & 1   \\[-1.70ex]
3            & 1                                          & 0.2511194029812406                & -0.05111940298124052              & 1   \\[-1.70ex]
4            & 1                                          & 0.2511194029812406                & -0.05111940298124052              & 1  \\[-1.70ex]
5            & 1                                          & 0.1488805970187594                 & 0.05111940298124053              & 1 \\[-1.70ex]
6            & 1                                          & 0.1488805970187594                 & 0.05111940298124052              & 1 \\[-1.70ex]
7            & 1                                          & 0.2511194029812406                  & 0.05111940298124052              & 1   \\[-1.70ex]
8           &  1                                          & 0.2511194029812406                  & 0.05111940298124052              & 1   \\
\hline
Average & 1                                          & 0.2000000000000000                 & 0                                                 & 1 \\
\hline
\end{tabular}
\end{center}
\label{table:1}
\end{table}
\end{subtables}

\begin{subtables}\label{ExTrac}
\begin{table}[H]\footnotesize
\begin{center}  
\caption{Components of stress tensor for simple extension mode under uniform traction boundary condition.} 
\begin{tabular}{l*{7}{c}r} \\[-1.70ex]
\hline
Number             & $P_{11}$(GPa)  & $P_{12}$(GPa)  & $P_{21}$(GPa)  & $P_{22}$(GPa)   \\
\hline
1             & 1.039825026165646    &  0.04820528954458368    & 0.04866128855931545        & -0.2416297163702212  \\[-1.70ex]
2             & 1.320203739079774    & -0.03982502616563410   & -0.03874395172276655        & -0.04866128855930084    \\[-1.70ex]
3             & 0.6802710990595731  &  0.02932516104756559    & 0.02946016137955194         & -0.1912627316990225   \\[-1.70ex]
4             & 0.9597001356950392  & -0.04029986430498470   & -0.03937749821610080        & -0.03937749821606967  \\[-1.70ex]
5             & 1.034848686668402    &  0.03484868666840837    & 0.03433176606052552         &  0.03433176606052036  \\[-1.70ex]
6             & 1.300707598404531    & -0.04322895004735811   & -0.04357035273468100         &  0.2559592388690298 \\[-1.70ex]
7             & 0.7026611918618899  &  0.03821747693482153     & 0.03716704160995366         &  0.02792845493580790   \\[-1.70ex]
8             & 0.9617825230651917  & -0.02724277367740260   & -0.02792845493579822          &  0.2027117749793306   \\
\hline
Average  & 1.000000000000006   & -0.00000000000000004    & 0.000000000000000001         & 0.0000000000000093 \\
\hline
\end{tabular}
\end{center}
\label{table:1}
\end{table}

\begin{table}[H]\footnotesize
\begin{center}  
\caption{Components of deformation gradient for simple extension mode under uniform traction boundary condition.} 
\begin{tabular}{l*{7}{c}r} \\[-1.70ex]
\hline
Number             & $F_{11}$  & $F_{12}$  & $F_{21}$  & $F_{22}$   \\
\hline
1            & 1.004937304707261     & -2.689342182920912E-4      & 8.747038856957929E-4              & 0.9969235353724314  \\[-1.70ex]
2            & 1.005765870068560     & -1.097499579591548E-3      & 6.054301474451633E-4              & 0.9971928091106821     \\[-1.70ex]
3            & 1.005765870068560      & 5.280152905005938E-5       & 6.054301474451633E-4             & 0.9960071216113445   \\[-1.70ex]
4            & 1.007269717597589     & -1.451045999979014E-3       & 5.567276539679379E-4             & 0.9960558241048217   \\[-1.70ex]
5            & 1.007549975212814      & 8.729795948439010E-4      & -9.583423724183159E-5             & 0.9963105708813713  \\[-1.70ex]
6            & 1.008691889025950     & -2.689342182920912E-4      & -7.087987283018901E-4             & 0.9969235353724314 \\[-1.70ex]
7            & 1.008691889025950      & 2.172935797205065E-3      & -7.087987283018901E-4             & 0.9955870601989383   \\[-1.70ex]
8            & 1.010812023294105      & 5.280152905005938E-5      & -1.128860140708145E-3             & 0.9960071216113445   \\
\hline
\end{tabular}
\end{center}
\label{table:1}
\end{table}
\end{subtables}

\begin{subtables}\label{ShTrac}
\begin{table}[H]\footnotesize
\begin{center}  
\caption{Components of stress tensor for simple shear mode under uniform traction boundary condition.} 
\begin{tabular}{l*{7}{c}r} \\[-1.70ex]
\hline
Number             & $P_{11}$(GPa)  & $P_{12}$(GPa)  & $P_{21}$(GPa)  & $P_{22}$(GPa)   \\
\hline
1             &  0.04560639290814648    & 0.9543878256031890       & 0.9543936070918476       & -0.015047074289714631  \\[-1.70ex]
2             & -0.01504707428965104    & 0.9543936070918378       & 0.9543878256031800        & 0.045606392908200080    \\[-1.70ex]
3             & -0.06074680125388290    & 1.060613153272475         & 1.061397992038798        &  -0.068434256720068380   \\[-1.70ex]
4             &  0.03018748263543562    & 1.030187482635433         & 1.029820575266174          & 0.029820575266171635   \\[-1.70ex]
5             &  0.02982057526613236    & 1.029820575266177         & 1.030187482635437          & 0.030187482635397278  \\[-1.70ex]
6             & -0.06843425672007669    & 1.061397992038797         & 1.060613153272471         & -0.060746801253881974 \\[-1.70ex]
7             & -6.77692342451637E-3    & 0.9546093951216100       & 0.9545899689704889        & 0.045390604878376570   \\[-1.70ex]
8             &  0.04539060487842260    & 0.9545899689704804       & 0.9546093951216026       & -6.7769234244772734E-3   \\
\hline
Average         & 0.0000000000000013  & 1.000000000000000    & 1.000000000000000     & 0.0000000000000004 \\
\hline
\end{tabular}
\end{center}
\label{table:1}
\end{table}

\begin{table}[H]\footnotesize
\begin{center}  
\caption{Components of deformation gradient for simple shear mode under uniform traction boundary condition.} 
\begin{tabular}{l*{7}{c}r} \\[-1.70ex]
\hline
Number             & $F_{11}$  & $F_{12}$  & $F_{21}$  & $F_{22}$   \\
\hline
1            & 1.000242207581661      & 5.764948592700328E-3     & 6.143346763082536E-3              & 0.9998638094113029  \\[-1.70ex]
2            & 0.9998638094113029    & 6.143346763057936E-3     & 5.764948592724816E-3              & 1.000242207581661   \\[-1.70ex]
3            & 0.9998638094113029    & 0.01791376094498578       & 5.764948592724816E-3              & 0.9997780280404228   \\[-1.70ex]
4            & 1.000156756804026     & 0.01762081355226269        & 5.390318334383442E-3              & 1.000152658298764   \\[-1.70ex]
5            & 1.000152658298764     & 5.390318334358900E-3      & 0.01762081355228730                & 1.000156756804026   \\[-1.70ex]
6            & 0.9997780280404227   & 5.764948592700328E-3      & 0.01791376094501020                & 0.9998638094113029   \\[-1.70ex]
7            & 0.9997780280404227   & 0.01892022363786956       & 0.01791376094501020                 & 1.000784490733306   \\[-1.70ex]
8            & 1.000784490733307     & 0.01791376094498578       & 0.01892022363789369                 & 0.9997780280404228   \\
\hline
\end{tabular}
\end{center}
\label{table:1}
\end{table}
\end{subtables}



\begin{thebibliography}{47}
\providecommand{\natexlab}[1]{#1}
\providecommand{\url}[1]{\texttt{#1}}
\expandafter\ifx\csname urlstyle\endcsname\relax
  \providecommand{\doi}[1]{doi: #1}\else
  \providecommand{\doi}{doi: \begingroup \urlstyle{rm}\Url}\fi


\bibitem[{Blanco et~al.(2014)Blanco, S{\'a}nchez, de~Souza~Neto, and
  Feij{\'o}o}]{blanco2014variational}
Blanco, P.~J., S{\'a}nchez, P.~J., de~Souza~Neto, E.~A., Feij{\'o}o, R.~A.,
  2014. Variational foundations and generalized unified theory of rve-based
  multiscale models. Archives of Computational Methods in Engineering, 1--63.

\bibitem[{de~Souza~Neto et~al.(2015)de~Souza~Neto, Blanco, S{\'a}nchez, and
  Feij{\'o}o}]{de2015rve}
de~Souza~Neto, E., Blanco, P., S{\'a}nchez, P., Feij{\'o}o, R., 2015. An
  rve-based multiscale theory of solids with micro-scale inertia and body force
  effects. Mechanics of Materials 80, 136--144.

\bibitem[{Feyel and Chaboche(2000)}]{feyel2000fe}
Feyel, F., Chaboche, J.-L., 2000. Fe 2 multiscale approach for modelling the
  elastoviscoplastic behaviour of long fibre sic/ti composite materials.
  Computer methods in applied mechanics and engineering 183~(3), 309--330.

\bibitem[{Hill(1963)}]{hill1963elastic}
Hill, R., 1963. Elastic properties of reinforced solids: some theoretical
  principles. Journal of the Mechanics and Physics of Solids 11~(5), 357--372.

\bibitem[{Hill(1967)}]{hill1967essential}
Hill, R., 1967. The essential structure of constitutive laws for metal
  composites and polycrystals. Journal of the Mechanics and Physics of Solids
  15~(2), 79--95.

\bibitem[{Hill(1972)}]{hill1972constitutive}
Hill, R., 1972. On constitutive macro-variables for heterogeneous solids at
  finite strain. Proceedings of the Royal Society of London. A. Mathematical
  and Physical Sciences 326~(1565), 131--147.

\bibitem[{Hori and Nemat-Nasser(1999)}]{hori1999two}
Hori, M., Nemat-Nasser, S., 1999. On two micromechanics theories for
  determining micro--macro relations in heterogeneous solids. Mechanics of
  Materials 31~(10), 667--682.

\bibitem[{Hughes(2012)}]{hughes2012finite}
Hughes, T.~J., 2012. The finite element method: linear static and dynamic
  finite element analysis. Courier Dover Publications.

\bibitem[{Mandel(1972)}]{mandel1972plasticite}
Mandel, J., 1972. Plasticit{\'e} classique et viscoplasticit{\'e}: CISM-1971.
  Udine, Springer-Verlag, Vienna, New York.

\bibitem[{Miehe et~al.(2002)Miehe, Schotte, and
  Lambrecht}]{miehe2002homogenization}
Miehe, C., Schotte, J., Lambrecht, M., 2002. Homogenization of inelastic solid
  materials at finite strains based on incremental minimization principles.
  application to the texture analysis of polycrystals. Journal of the Mechanics
  and Physics of Solids 50~(10), 2123--2167.

\bibitem[{Miehe et~al.(1999)Miehe, Schr{\"o}der, and
  Schotte}]{miehe1999computational}
Miehe, C., Schr{\"o}der, J., Schotte, J., 1999. Computational homogenization
  analysis in finite plasticity simulation of texture development in
  polycrystalline materials. Computer Methods in Applied Mechanics and
  Engineering 171~(3), 387--418.

\bibitem[{Molinari and Mercier(2001)}]{molinari2001micromechanical}
Molinari, A., Mercier, S., 2001. Micromechanical modelling of porous materials
  under dynamic loading. Journal of the Mechanics and Physics of Solids 49~(7),
  1497--1516.

\bibitem[{Nemat-Nasser(1999)}]{nemat1999averaging}
Nemat-Nasser, S., 1999. Averaging theorems in finite deformation plasticity.
  Mechanics of Materials 31~(8), 493--523.

\bibitem[{Nemat-Nasser and Hori(2013)}]{nemat2013micromechanics}
Nemat-Nasser, S., Hori, M., 2013. Micromechanics: overall properties of
  heterogeneous materials. Elsevier.

\bibitem[{Nguyen et~al.(2011)Nguyen, Lloberas-Valls, Stroeven, and
  Sluys}]{nguyen2011homogenization}
Nguyen, V.~P., Lloberas-Valls, O., Stroeven, M., Sluys, L.~J., 2011.
  Homogenization-based multiscale crack modelling: from micro-diffusive damage
  to macro-cracks. Computer Methods in Applied Mechanics and Engineering
  200~(9), 1220--1236.

\bibitem[{Peri{\'c} et~al.(2011)Peri{\'c}, de~Souza~Neto, Feij{\'o}o, Partovi,
  and Molina}]{peric2011micro}
Peri{\'c}, D., de~Souza~Neto, E., Feij{\'o}o, R., Partovi, M., Molina, A.,
  2011. On micro-to-macro transitions for multi-scale analysis of non-linear
  heterogeneous materials: unified variational basis and finite element
  implementation. International Journal for Numerical Methods in Engineering
  87~(1-5), 149--170.

\bibitem[{Pham et~al.(2013)Pham, Kouznetsova, and Geers}]{pham2013transient}
Pham, K., Kouznetsova, V., Geers, M., 2013. Transient computational
  homogenization for heterogeneous materials under dynamic excitation. Journal
  of the Mechanics and Physics of Solids 61~(11), 2125--2146.

\bibitem[{Reina(2011)}]{reina2011multiscale}
Reina, C., 2011. Multiscale modeling and simulation of damage by void
  nucleation and growth. Ph.D. thesis, California Institute of Technology.

\bibitem[{Reina et~al.(2013)Reina, Li, Weinberg, and
  Ortiz}]{reina2013micromechanical}
Reina, C., Li, B., Weinberg, K., Ortiz, M., 2013. A micromechanical model of
  distributed damage due to void growth in general materials and under general
  deformation histories. International Journal for Numerical Methods in
  Engineering 93~(6), 575--611.

\bibitem[{Ricker et~al.(2009)Ricker, Mergheim, and
  Steinmann}]{ricker2009multiscale}
Ricker, S., Mergheim, J., Steinmann, P., 2009. On the multiscale computation of
  defect driving forces. International Journal for Multiscale Computational
  Engineering 7~(5).

\bibitem[{Smit et~al.(1998)Smit, Brekelmans, and Meijer}]{smit1998prediction}
Smit, R., Brekelmans, W., Meijer, H., 1998. Prediction of the mechanical
  behavior of nonlinear heterogeneous systems by multi-level finite element
  modeling. Computer Methods in Applied Mechanics and Engineering 155~(1),
  181--192.

\bibitem[{Suquet(1987)}]{suquet1987elements}
Suquet, P., 1987. Elements of homogenization for inelastic solid mechanics.
  Homogenization techniques for composite media 272, 193--278.

\bibitem[{Terada et~al.(2000)Terada, Hori, Kyoya, and
  Kikuchi}]{terada2000simulation}
Terada, K., Hori, M., Kyoya, T., Kikuchi, N., 2000. Simulation of the
  multi-scale convergence in computational homogenization approaches.
  International Journal of Solids and Structures 37~(16), 2285--2311.

\bibitem[{Willis(1981)}]{willis1981variational}
Willis, J.~R., 1981. Variational and related methods for the overall properties
  of composites. Advances in applied mechanics 21, 1--78.

\bibitem[{Zohdi and Wriggers(2008)}]{zohdi2008introduction}
Zohdi, T.~I., Wriggers, P., 2008. An introduction to computational
  micromechanics. Springer Science \& Business Media.

\end{thebibliography}
\end{document}